\documentclass[aps,pre,showpacs,showkeys,twocolumn,superscriptaddress,
floats,floatfix,preprintnumbers,longbibliography]{revtex4-1}
\usepackage{amssymb,amsmath}
\usepackage{bm,url,graphics,subfigure,epsfig,rotating,float}
\usepackage{mathrsfs} 
\usepackage{color,soul}
\graphicspath{{./}{figs/}}
\usepackage{braket}
\usepackage{hyperref}
\usepackage[normalem]{ulem}   
\usepackage[dvipsnames, table, xcdraw]{xcolor}
\usepackage{relsize}
\usepackage{slashbox}
\def \WM    {W_{\rm Met}}
\def \WG    {W_{\rm Gla}}
\def \Vref	{V_{\rm ref}}
\def \mY    {\mathcal{Y}}
\def \ft    {\tilde{f}}

\def \dtxx     {{\rm d}^2\xx}
\def \Pnot  {P_{\rm 0}}
\def \Pc  {P_{\rm c}}

\def \qstar {q_{\ast}}
\def \Geff {G_{\rm eff}}
\def \fhat {\hat{f}}
\def \kB   {k_{\rm B}}

\def \Gone {G_{\rm 1}}
\def \Gzero {G_{\rm 0}}
\def \PT    {\mathcal{P}}

\def \ET    {\mbox{ET}}

\def \bET 	{\textbf{\ET}}

\def \qq     {\bm{q}}

\def \aa {a}

\def \mP {\mathcal{P}}

\def  \xx  {{\bm x}}
\def  \XX  {{\bm X}}

\def \grad {{\bm \nabla}}

\def \lap {\nabla^2}

\def \fhat {\hat{f}}
\newcommand{\avg}[1]{\left\langle #1\right\rangle}

\def \Fv  {\mbox{FvK}}

\def \bA {\textbf{A}}

\newcommand{\eq}[1]{~(\ref{#1})}
\newcommand{\Eq}[1]{Eq.~(\ref{#1})}
\newcommand{\Fig}[1]{Fig.~(\ref{#1})}
\newcommand{\subfig}[2]{Fig.~(\ref{#1}#2)}

\newcommand{\bfig}{\begin{figure}}
\newcommand{\efig}{\end{figure}}
\newcommand{\bc}{\begin{center}}
\newcommand{\ec}{\end{center}}
\newcommand{\bea}{\begin{eqnarray}}
\newcommand{\eea}{\end{eqnarray}}

\def \FvK {\mbox{FvK}}

\def \Es    {E_{\rm stretch}}
\def \Eb    {E_{\rm bend}}
\def \flm   {\tilde{f}_{\ell,m}}
\def \Ylm   {\mathcal{Y}^m_{\ell}}
\def \Sum {\mathlarger{\sum}}
\def \SUM {\mathlarger{\mathlarger{\sum}}}

\def \Xij {X_{\rm ij}}
\def \Xijm {X_{\rm ij-1}}
\def \XXi 	{\XX_{\rm i}}
\def \XXj 	{\XX_{\rm j}}
\def  \XXij  {\XX_{\rm ij}}
\def  \XXijm  {\XX_{\rm ij-1}}
\def \lzij  {\ell^{\rm 0}_{\rm ij}}
\def \nhat {\hat{{\bm n}}}

\def \cAib {{\mathcal A}_i}
\def \LL {{\bm L}}
\def \aij {\alpha_{\rm ij}}
\def \bij {\beta_{\rm ij}}
\def \bijm {\beta_{ij-1}}

\def \bijm 	{\beta_{ij-1}}
\def \cAjib	{{\mathcal A}_{j(i)}}
\def \etal 	{et al\/.}

\def \FVK {F\"oppl--von K\'arm\'an }
\begin{document}
\title{Active buckling of pressurized
  spherical shells : Monte Carlo Simulation}
\author{Vipin Agrawal}
\email{vipin.agrawal@su.se}
\affiliation{ Nordita, KTH Royal Institute of Technology and
Stockholm University, Roslagstullsbacken 23, 10691 Stockholm, Sweden}
\affiliation{ Department of Physics, Stockholm university, Stockholm,Sweden.}
\author{Vikash Pandey}
\email{vikash.pandey@su.se}
\affiliation{ Nordita, KTH Royal Institute of Technology and
Stockholm University, Roslagstullsbacken 23, 10691 Stockholm, Sweden}
\author{Dhrubaditya Mitra}
\email{dhruba.mitra@gmail.com}
\affiliation{ Nordita, KTH Royal Institute of Technology and
Stockholm University, Roslagstullsbacken 23, 10691 Stockholm, Sweden}
\date{\today}
\begin{abstract}
We study the buckling of pressurized spherical shells by 
Monte Carlo simulations in which the
detailed balance is explicitly broken -- thereby driving the shell
active, out of thermal equilibrium.
Such a shell typically has  either higher (active) or
lower (sedate) fluctuations
compared to one in thermal equilibrium depending on how the detailed balance
is broken.
We show that, for the same set of elastic parameters, a shell
that is not buckled in thermal equilibrium can be buckled if turned
active. Similarly a shell that is buckled in thermal equilibrium
can unbuckle if sedated.
Based on this result, we suggest that it is possible to experimentally design
microscopic elastic shells whose buckling can be optically controlled.
\end{abstract}
\maketitle
Thin spherical shells are commonly found in many natural and engineering
settings. Their sizes can vary over a very large range --  from hundred
meters, e.g., the Avicii Arena Stockholm ~\footnote{
  There are several geodesic domes with sizes ranging from 10 to 200
  meters. } down to about hundred nanometers, e.g.,
viral capsules~\cite{buenemann2008elastic,buenemann2008elastic,
  michel2006nanoindentation} and
exosomes~\cite{pegtel2019exosomes,cavallaro2019label}.
The elastic properties of shells, including conditions under which buckling
can occur, have been extensively studied~\cite{Love2013treatise,LLelast,
  koiter1963progress,koiter1976buckling,pogorelov1988bendings,
  hutchinson2016buckling}. 
Interest in this traditional field of applied mathematics has been
rekindled in the past decades because of possible applications to
biology and nanoscience~\cite{vliegenthart2006mechanical,gao2001elasticity,
  gordon2004self, buenemann2008elastic, vella2012indentation,
  vorselen2017competition,wu2018comparison,
  vorselen2020mechanical,cavallaro2021multiparametric,
  michel2006nanoindentation,zoldesi2008elastic}.
For example, the elastic shell is used as model for nuclear
membrane~\cite{jackson2022dynamics}.
Furthermore,  the cell membrane, although often modeled simply
as a fluid membrane, is dynamically tethered to the cytoskeleton
-- therby acquiring effective in-plane elastic properties. 
For example,  it has been shown~\cite{hw2002stomatocyte} that to capture the
 stomatocyte--discocyte--echinocyte
  sequence of the human red blood cell within one unified model it
  is necessary to introduce nonlinear in-plane shear elastic modulus of
  the membrane.  Numerical simulations of flowing RBCs that faithfully
  reproduce experimental observations also must use nonlinear shear elastic
  modulus~\cite{fedosov2010multiscale, fedosov2010systematic,
    Fedosov2010multi, fedosov2011predicting, freund2014numerical,
    zhu2014microfluidic}.
Crucially, it has been  shown that for small enough shells
the thermal fluctuations
can bring down the critical buckling pressure by a large
amount~\cite{paulose2012fluctuating,kovsmrlj2017statistical}.
This opens up the intriguing possibility of how
the elastic properties of shells, in particular buckling,
will change if they are turned active -- driven out of
thermal equilibrium.
\begin{figure}
  \includegraphics[width=\columnwidth]{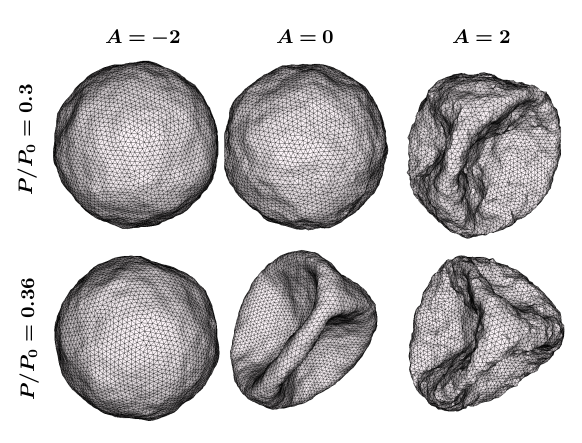}
\caption{\label{fig:buck} {\bf Active buckling:}
  Typical snapshots from our simulations for activity $A = -2$, $0$ and $2$
  (from left to right), and pressure $P = 0.30\Pnot$ (top row) and
  $0.36\Pnot$ (bottom row) where $\Pnot$ is the critical buckling
  pressure obtained from the mechanical theory of elastic shells, i.e.,
  at zero temperature. We use  $N = 5120$, $\FvK = 4616$, and
  $\ET = 8$. The middle column, $A = 0$, corresponds to
  shells in thermal equilibrium -- an unbuckled shell
  buckles upon increasing $P/\Pnot$
  from $0.30$ to $0.36$. This is consistent with the
  results of Refs.~\cite{paulose2012fluctuating,kovsmrlj2017statistical}.
  Top row: As activity is increased to $2$ (right column) the shell buckles.
  Bottom row: Whereas as activity is decreased to $-2$ (left column) the
   shell, that was buckled in thermal equilibrium, does not buckle 
   at the same pressure. 
} 
\end{figure}
The fundamental property of living
matter is that they are not in thermal
equilibrium~\cite{schrodinger2012life}
even when they are statistically stationary.
They are active -- 
they consume energy and generate entropy~\cite{gnesotto2018broken}.
The statistical and mechanical properties of active
matter is a current topic of considerable
interest~\cite{marchetti2013hydrodynamics,ramaswamy2010mechanics}.
The fluctuations of  the membrane of living cells 
have active components, in addition to the thermal fluctuations, due
to, e.g., driving by the  active
cytoskeleton~\cite{peng2013lipid,turlier2016equilibrium,biswas2017mapping,
  turlier2019unveiling,manikandan2022estimate}.
Active shells can also be synthetically
designed, e.g., by embedding 
certain proteins, who acts as active pumps when irradiated with light of
certain frequencey, in bi-lipid membranes~\cite{girard2005passive,
  manneville2001active}.
Shells made out of hard--magnetic elastomers can be turned active
by an external magnetic field~\cite{yan2021magneto}. 

In this paper, we study the buckling of pressurized active spherical shells 
using the Monte Carlo (MC) simulations~\cite{kumar2020nonequilibrium}
in which detailed balance is explicitly broken
-- thereby driving the shell active, out of thermal equilibrium.
Such a shell typically has either higher or lower fluctuations than
thermal ones depending on how the detailed balance is broken. 
We call such non-equilibrium stationary states \textit{active}
and \textit{sedate} respectively.
We show that, within the right range of elastic parameters, a shell
that is not buckled in thermal equilibrium can be buckled if turned
active. Similarly a shell that is buckled in thermal equilibrium
can unbuckle if turned sedate, see \Fig{fig:buck}.
Based on our study, we suggest that it is possible to experimentally design
microscopic elastic shells whose buckling can be optically controlled. 

Let us briefly summarize, following Refs.~\cite{paulose2012fluctuating,
kovsmrlj2017statistical} the model and the key results of theory of
thin elastic shells in \textit{thermal equilibrium}.
A pressurized elastic shell is described by
an effective  Hamiltonian, $\Geff = \Gzero + \Gone$ where,
\begin{subequations}
  \begin{align}
    \Gzero[f] &= \frac{1}{2}\int \dtxx  \left[ B (\lap f)^2
    - \frac{PR}{2}\mid \grad f\mid^2 + \frac{Y}{R^2}f^2\right], {~\text {and}}  \\
    \Gone[f] &= \frac{Y}{2}
    \int \dtxx \left[\left(\frac{1}{2}\PT_{ij}\partial_i f \partial_j f\right)^2
      - \frac{f}{R}\PT_{ij}\partial_i f \partial_j f\right]\/.
  \end{align}
\end{subequations}
Here $\xx = (x_1,x_2)$ is a two-dimensional Cartesian coordinate system
and
$\PT_{ij} \equiv \delta_{ij} - \partial_i\partial_j/\lap$ 
is the transverse projection operator. 
The out-of-plane displacement is $h(\bm x) = f_0(\bm x) + f(\bm x)$ where $f_0(\bm x)$ is the
uniform contraction of the sphere in response to the external pressure.
The difference between the external and the internal pressure is $P$.
The part $\Gzero$ is harmonic and the part $\Gone$ is anharmonic.
In this model, we assume the shell to be  amorphous and homogeneous with
radius $R$, bending modulus $B$ and (two--dimensional) Young's modulus $Y$.
Two non-dimensional numbers determine the elastic behavior
of such shells, the F\"oppl--von-Karman number
and the Elasto-thermal number, defined respectively as
\begin{equation}
  \Fv \equiv \frac{YR^2}{B}\/,\quad \ET \equiv \frac{\kB T}{B}\sqrt{\Fv}\/,
  \label{eq:ndim}
\end{equation}
where $\kB$ is the Boltzmann constant and $T$ is temperature. 
At constant $\ET$, the effects of anharmonicity increases with 
$\Fv$ whereas at constant elastic modulii the
effects of thermal fluctuations increases with $\ET$.
Ignoring the anharmonic contribution,
using standard tools of equilibrium statistical mechanics
it is straightforward~\cite[][Eq. 4]{paulose2012fluctuating} to calculate
the spectrum of fluctuations
\begin{equation}
  S(\qq) \equiv \avg{\fhat(\qq)\fhat(-\qq)}
  = \frac{\kB T}{a\left(Bq^4 - \frac{PRq^2}{2} +\frac{Y}{R^2}\right)}\/,
  \label{eq:S}
\end{equation}
where $\fhat(\qq)$ is the Fourier transform of $f(\xx)$ and
$a$ is the area of integration in the $(x_1,x_2)$ plane.
In equilibrium, the symbol $\avg{\cdot}$ denotes thermal averaging;
whereas for active cases, it denotes averaging over the non--equilibrium
stationary states.
Note that $S(\qq)$ blows-up for
\begin{subequations}
  \begin{align}
   P &= \Pnot \equiv \frac{4B}{R}\qstar^2\/,\quad \text{where} \\
  \qstar &\equiv \left(\frac{Y}{B R^2} \right)^{1/4} = \frac{\Fv^{1/4}}{R}\/,
  \label{eq:Pnot}
  \end{align}
\end{subequations}
where $\Pnot$ is the buckling pressure,
independent of temperature, obtained within the traditional
theory~\cite{hutchinson2016buckling} of buckling of pressurized shells.
For a large F\"oppl--von-Karman number, 
$\qstar > 1/R$
is the buckling mode.
Refs.~\cite{paulose2012fluctuating,kovsmrlj2017statistical} used
renormalization group (RG) techniques to show that the effects of the
anharnomic terms is to renormalize the parameters appearing in the
bare theory, i.e.,  $P$, $B$, and $Y$ in \eq{eq:S} must be
replaced by their scale--dependent, renormalized
versions, see Ref.~\cite[][Eq. 18]{kovsmrlj2017statistical}.
Consequently both the pressure and the critical buckling pressure
are renormalized and buckling is obtained if  both of
these quantities are
equal for  a length scale which must be smaller than the radius of the
sphere~\cite{kovsmrlj2017statistical}. 
The results of this RG analysis were validated by Monte Carlo
simulations of spherical shell, randomly triangulated with $N$ grid points,
with discretized bending and stretching energies
that translate directly into a macroscopic elastic
modulii~\cite{paulose2012fluctuating,gompper2004triangulated,
itzykson1986proceedings}.
Our Monte Carlo code, described in detail in Ref.~\cite{agrawal2022memc},
    closely follows that of Ref.~\cite{paulose2012fluctuating}, and
faithfully reproduces these results.
We  incorporate activity into this model in the
following manner. 

Over the years, many theoretical models~\cite{ramaswamy2000nonequilibrium,
  rao2001active, loubet2012effective,maitra2014activating, hawkins2014stress,
  yin2021bio, goriely2017five}, have been suggested to incorporate the
effects of active fluctuations into models of membranes.
We use a method that is well suited to use the Monte Carlo
setup and has been used before to study Ising models
out of equilibrium~\cite{kunsch1984non,godreche2009nonequilibrium,
  godreche2013rates,kumar2020nonequilibrium}
-- the idea is to break detailed balance while preserving
stationarity. 
In equilibrium Monte Carlo simulations
two common choices of the  transition rate
from one state to another
are the  Metropolis ($\WM$) and the Glauber ($\WG$),
given respectively by,
\begin{subequations}
  \begin{align}
  \WM &= {\rm min}\left[1,\exp\left(-\frac{E}{\kB T}\right)\right],\/\text{and}
\label{eq:WM}\\
  \WG &= \frac{1}{2}\left[1-\tanh\left(\frac{E}{2\kB T}\right)\right]\/,
  \label{eq:WG}
  \end{align}
  \label{eq:W}
\end{subequations}
where $\kB$ is the Boltzmann constant, $T$ is the temperature 
and $E$ is the difference in energy between the two states.
To drive the membrane out of equilibrium, following
Ref.~\cite{kumar2020nonequilibrium},
we replace $E$ by $E + \Delta E$ where $\Delta E$ is a constant.
This guarantees that detailed balance is broken and the
amount by which it is broken is $\Delta E$.
If $\Delta E$ is positive (negative) the probability of acceptance
of large fluctuations is decreased (increased).
Thus we define a dimensionless quantity
$A = - \Delta E/(\kB T)$ such that simulations with
positive $A$, \textit{active} simulations, have higher fluctuations than
equilibrium ones whereas for negative $A$, \textit{sedate}
simulations,
the fluctuations are less than the equilibrium ones.
For most of the simulations reported here we use the Metropolis
algorithm.
In some representative cases, for both equilibrium and
non-equilibrium simulations, we have checked that
both the Glauber and Metropolis algorithm gives the same result.

For lipid vesicles in thermal equilibrium,
standard techniques of
equilibrium statistical mechanics~\cite{turlier2019unveiling}
and  micropipet aspiration experiments show
$\Delta \alpha \propto (\kB T/4\pi B)\ln \sigma$
where $\Delta \alpha$ is the areal strain and
$\sigma$ is the surface tension.
For active membranes the same proportionality holds but the constant of
proportionality is different~\cite{manneville2001active}.
This experimental result was captured by the model in
Ref.~\cite{prost1996shape} which adds an additional
Ornstein--Uhlenbeck noise to the models of thermal membranes. 
Our active Monte Carlo scheme, in planar
membranes~\footnote{A. Fragkiadoulakis, S.K. Manikandan, and
  D. Mitra, unpublished.} reproduces the 
results of, Ref.~\cite{prost1996shape} and also 
the experimental result of Ref.~\cite{manneville2001active}. 

In summary, we incorporate the technique of
active Monte Carlo~\cite{kumar2020nonequilibrium}
into the Monte Carlo algorithm for spherical shells
in thermal equilibrium~\cite{paulose2012fluctuating,
  gompper2004triangulated,agrawal2022memc}
to simulate active shells. 
\begin{figure}
\centering
    \includegraphics[width=0.95\linewidth]{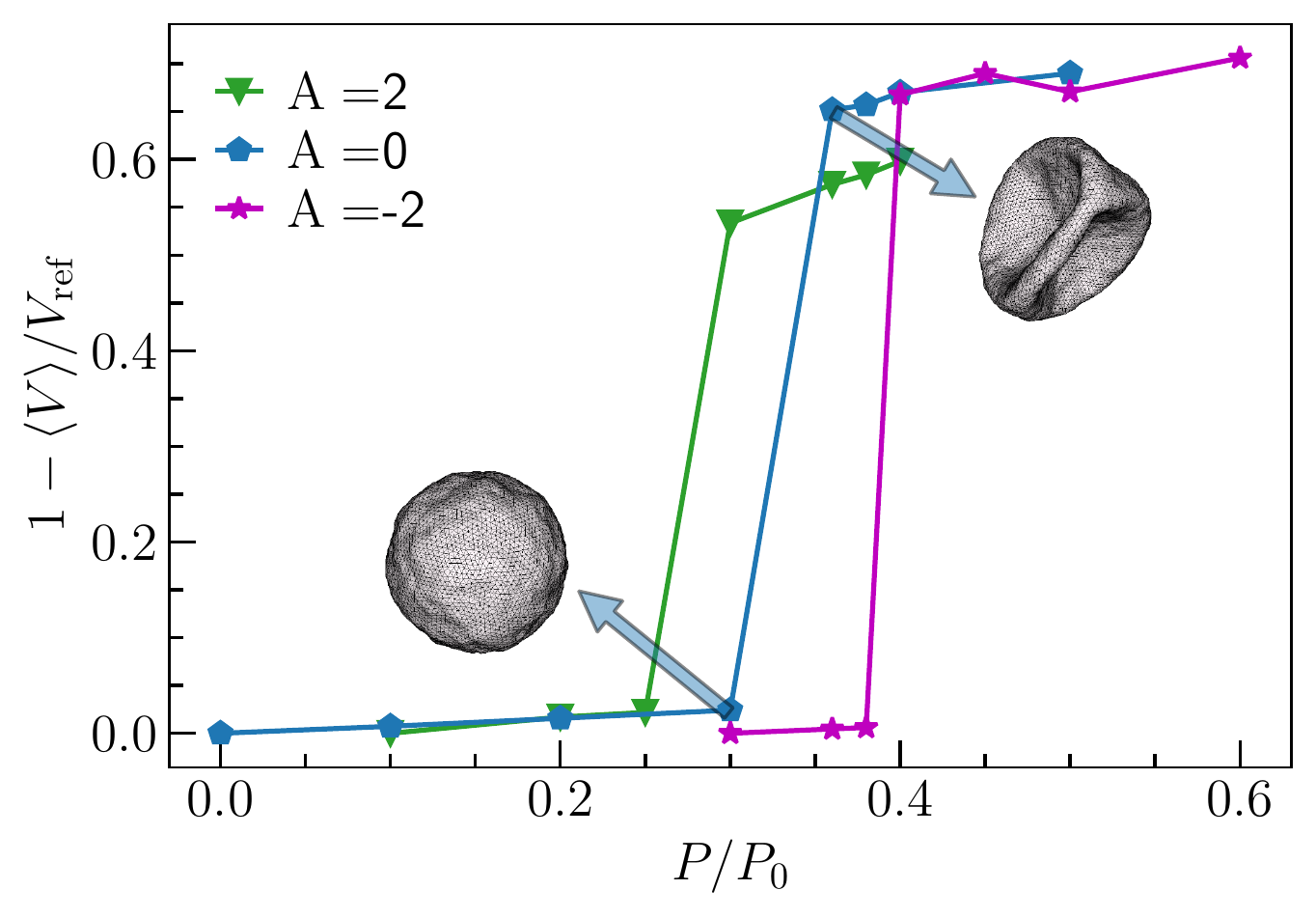}
    \caption{\label{fig:vol}{\bf Buckling under pressure:}
      Normalized change in volume as a function of external pressure
      for (blue) a shell in thermal equilibrium, (green) active ($A=2$),
      and (magenta) sedate ($A=-2$) for simulations with $\ET = 8 $ and
      $\FvK = 4616$ and number of grid points $N=5120$.
      Here $\avg{V}$ is the ensemble average of volume,
	  and $\Vref$ is the average volume at the smallest pressure difference.
      The error in $\avg{V}$ are the shades around the solid lines
      -- they are too small to be visible.
      The signature of buckling is the sudden
      large change in volume. The critical buckling pressure for the
      thermal case is consistent with
      Refs~\cite{paulose2012fluctuating,kovsmrlj2017statistical}.
      }
\end{figure}
\begin{figure}
\centering
    \includegraphics[width=0.9\columnwidth]{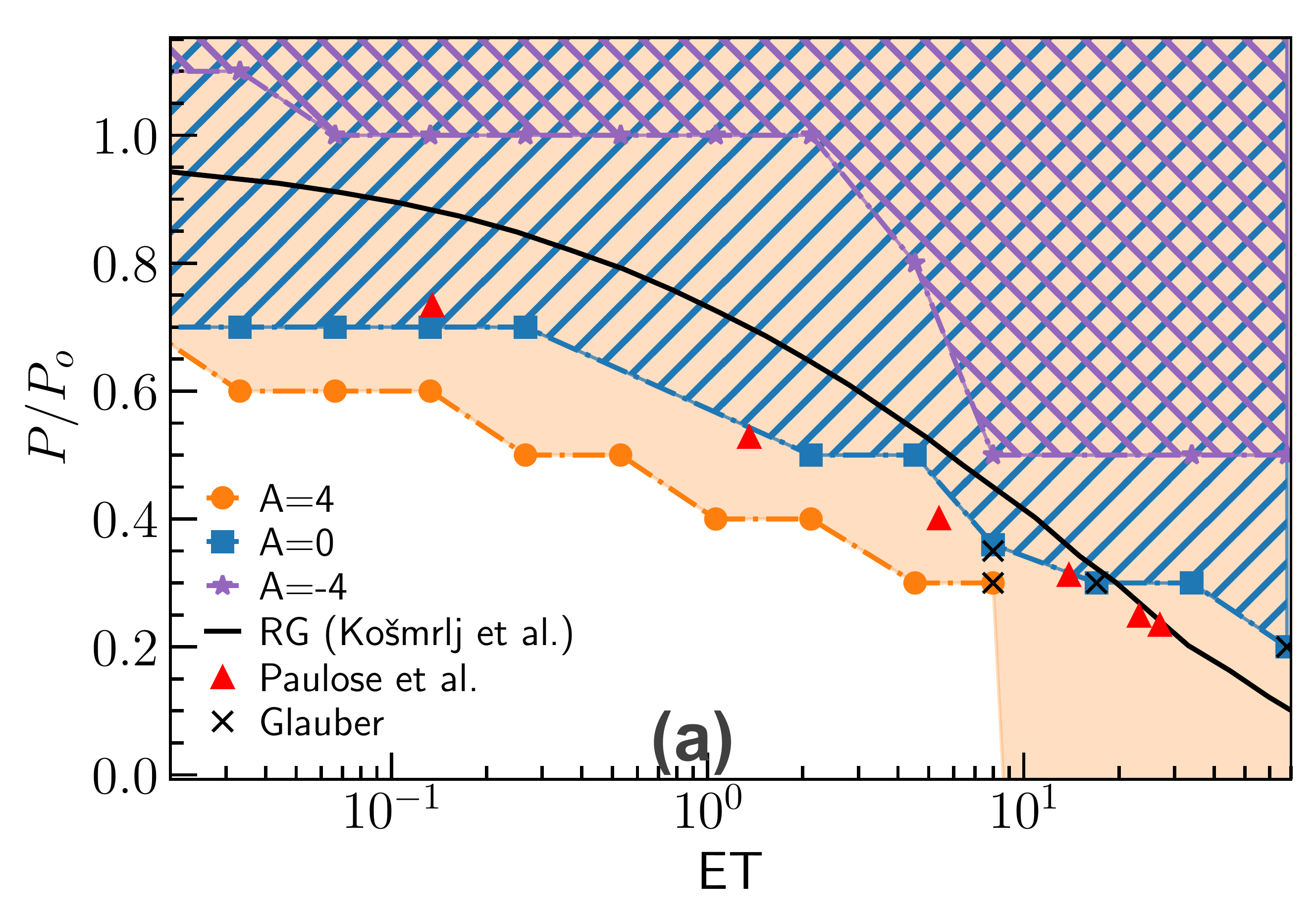}\\
    \includegraphics[width=0.9\columnwidth]{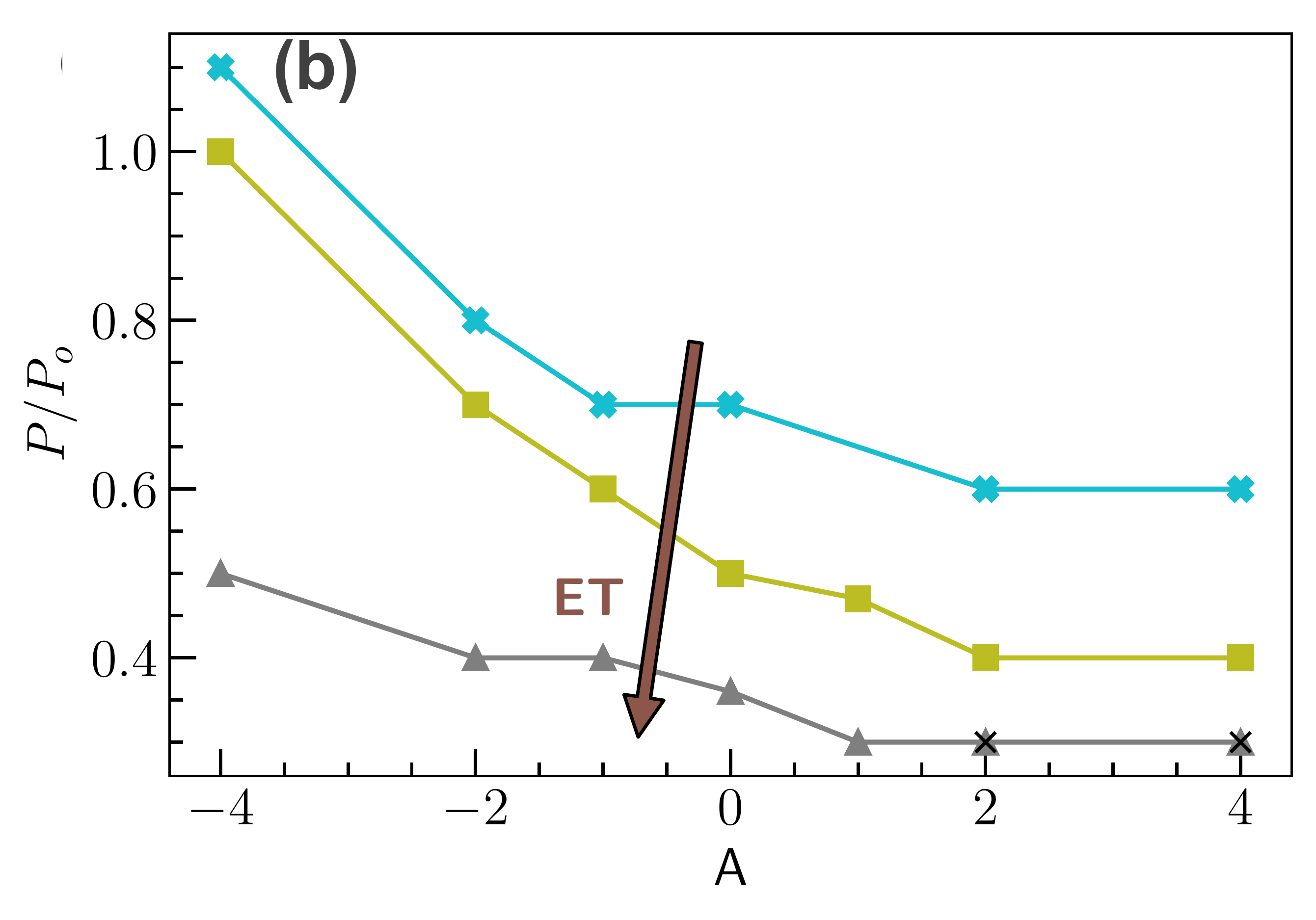}
    \caption{\label{fig:PD}{\bf Phase diagram}
The phase boundary in (a) the pressure--elasto--thermal
number plane for different activities and (b) in the
pressure--activity plane for different elasto--thermal
numbers (gray triangles for $\ET=7.99$, 
olive squares for $\ET=2.12$ 
and cyan cross for $\ET=0.03$).
In (a) the region where the buckled phase is obtained
in equilibrium is marked by blue lines.
The region where the buckled phase is obtained for $A=4$ is
shaded in light yellow. The region where the buckled phase
is obtained for $A=-4$ is marked by violet lines.
Cases marked with a cross uses the Glauber algorithm. 
In (a) the phase boundary obtained by RG
calculation~\cite{kovsmrlj2017statistical} is marked by a black line
and the simulation results by Ref.~\cite{paulose2012fluctuating} are
represented by red triangles.
}
\end{figure}

In \Fig{fig:vol} we show a typical plot of how the volume, $V$, of the spherical
shell changes as the external pressure is increased from a very small value.
The simulations are done in a constant pressure ensemble,
hence volume is a fluctuating quantity.
Henceforth, by volume we mean the average volume $\avg{V}$.
The average volume at the smallest pressure difference,
is the reference volume $\Vref$.
The error in $\avg{V}$, shown by the shaded regions in~\Fig{fig:vol}
are the variances -- they are too minute to be visible.
First consider the shell under thermal equilibrium.
Buckling shows up as a sharp decrease in volume accompanied by a typical
buckled shape, as shown in \Fig{fig:buck}.
The critical buckling pressure, $\Pc$, that we obtain is consistent with the
results of Refs.~\cite{paulose2012fluctuating,kovsmrlj2017statistical}.
We show the results of the simulations for both the active, $A=2$,
and the sedate, $A=-2$ cases. For the former the critical buckling
pressure decreases while for the latter the critical buckling pressure
increases.

Next we decompose the fluctuating height field, $f(\theta,\phi)$ in
spherical harmonics, $\mY_{\ell,m}(\theta,\phi)$:
\begin{subequations}
  \begin{align}
    f(\theta,\phi) &= \sum_{\ell,m}\ft_{\ell,m}\mY_{\ell}^m(\theta,\phi)\/, \quad
    \text{and define} \\
    S(\ell) &= \frac{4\pi}{(2\ell+1)\vert\ft_{00}\vert^2}\sum_{m=-\ell}^{\ell}\vert\ft_{\ell,m}\vert^2
  \end{align}
  \label{eq:spec}
\end{subequations}
In supplimentary material we compare typical plots of $S(\ell)$ 
for buckled and unbuckled shells.
Buckling is accompanied by appearence of a peak in $S(\ell)$
at a small $\ell$ value. For the equilibrium case, buckling as a function
of external pressure is an equilibrium phase transition
with the amplitude of the peak of $S(\ell)$ at small $\ell$ as the order
parameter~\cite{paulose2012fluctuating}.
But buckling at the fixed $P$ and $\ET$ as a function of activity is
not an equilibrium phase transition but can be considered as a dynamical
one. Nevertheless, we can still characterize buckling by appearance of
a peak in $S(\ell)$ for small $\ell$.

To obtain the phase diagram we use thirteen values of elasto--thermal number,
for each of which we use seven values of activity.
For a fixed choice of elasto--thermal number and activity we start our
simulations with an initial condition where the shell is 
a perfect sphere. 
Then we choose a fixed value of external pressure and run our simulations
till we reach a stationary state, which for zero activity is the
equilibrium state. Whether the shell is buckled or not is decided
by three checks: (a) significant decrease of volume (b) a peak at
small $\ell$ for $S(\ell)$ (c) visual inspection.
If the shell is not buckled we choose a higher external pressure and
start our simulations again from the same initial condition. 
The buckling pressure, $\Pc$ obtained for a set of parameters is given
in supplimentary material. 
This way we mark out the phase boundary in the pressure--elasto--thermal
number plane for different activities and in the pressure--activity plane for
different elasto--thermal numbers, see \Fig{fig:PD}.
In \subfig{fig:PD}{a} we also plot the phase boundary, obtained
through a RG calculation in Ref.~\cite{kovsmrlj2017statistical},
which agrees reasonably well with our numerical results for zero activity. 
Note that for large enough values of $\ET$ and $A$ we reach a part
of the phase diagram where the shell is unstable at zero
external pressure and can be made stable only with positive internal
pressure. This part of the phase diagram is not shown in \Fig{fig:PD}
although the relevant data are included in supplimentary material.
Note that at small $\ET$ for the sedate case it is possible to
have the shell remain unbuckled even for pressure higher $\Pnot$,
i.e., the shell is stabilized.

Several comments are now in order: One, most of our simulations
use $N = 5120$. We have repeated some of our
simulations with $N=20252$ and obtained the same buckling pressure.
Two, to obtain the buckling pressure we always start from the
same initial condition and imposed a fixed external pressure. 
Hence, the lines of phase separation we show, \Fig{fig:PD}, are not
continuous and will be improved if the phase diagram is sampled in
a finer resolution.
Three, experimentally, it is unclear how to implement the
sedate regime, negative $A$. 
Nevertheless, synthetic membranes that can be
turned active ($A>0$) optically, has been  already realized
by embedding  certain proteins in a bi-lipid membranes -- proteins that act
as active pumps when irradiated with light of certain
frequency~\cite{girard2005passive,manneville2001active}.
In such cases, only a fraction of points on the shell are active.
This can be incorporated in a straightforward manner in our code and
it would be interesting to see how the critical buckling pressure
changes as we change the fraction of active points. 
Four, bi--lipid membranes are semi-permeable~\cite{phillips2012physical}. 
As the shell buckles the fraction of solute increases, increasing the
partial pressure inside the shell. 
Experimentally, this can be avoided by using shell with holes in them.
We expect, in such cases the buckling pressure may change by a small amount. 
Five, as there are many different models of active elastic material
it behooves us to study the universallity of our result
by performing similar simulations in other models.
This is outside the scope of the present work. 

Finally, our simulations point towards the intriguing possibility that
within the right range of elastic parameters, a shell
that is not buckled in thermal equilibrium can be buckled if turned
optically  active. 
Based on this, we suggest that it is possible to experimentally design
microscopic elastic shells whose buckling can be optically controlled. 
In such devices it may be possible to drive flows at microscopic
scales by buckling and unbuckling of shells, optically.

\acknowledgements
We thank Apurba Dev, A. Fragkiadoulakis,
Sreekanth M. Manikandan, Pinaki Chaudhuri and
Bidisha Sinha for useful discussions.
We gratefully acknowledge the use of the following free
software packages: meshzoo~\cite{caroli2010robust,meshzoo},
matplotlib~\citep{Hunter:2007} and VisIt~\citep{HPV:VisIt}.
The simulations were performed on resources provided by
the Swedish National Infrastructure for Computing (SNIC) at PDC center for
high performance computing and  
the HPC facility supported by the Technical Division at the Department of
Physics, Stockholm University.
For the latter, we gratefully acknowledge generous help from
Mikica Kocic.
We acknowledge the financial support of the Swedish Research Council
through grants  638-2013-9243 and 2016-05225.
\section*{Code and data availability}
The source code used for the simulations of the study is freely available 
at \url{https://github.com/vipinagrawal25/MeMC/releases/tag/v1.1} ~\cite{agrawal2022memc}.
The simulation setup and the corresponding data are freely available on 
Zenodo with DOI: 10.5281/zenodo.6772570.
Python scripts are included with the data to generate all the figures.
\appendix
\section{Monte Carlo simulation of elastic shells in thermal equilibrium}
\label{sect:method}
\begin{figure}
  \includegraphics[width=0.8\columnwidth]{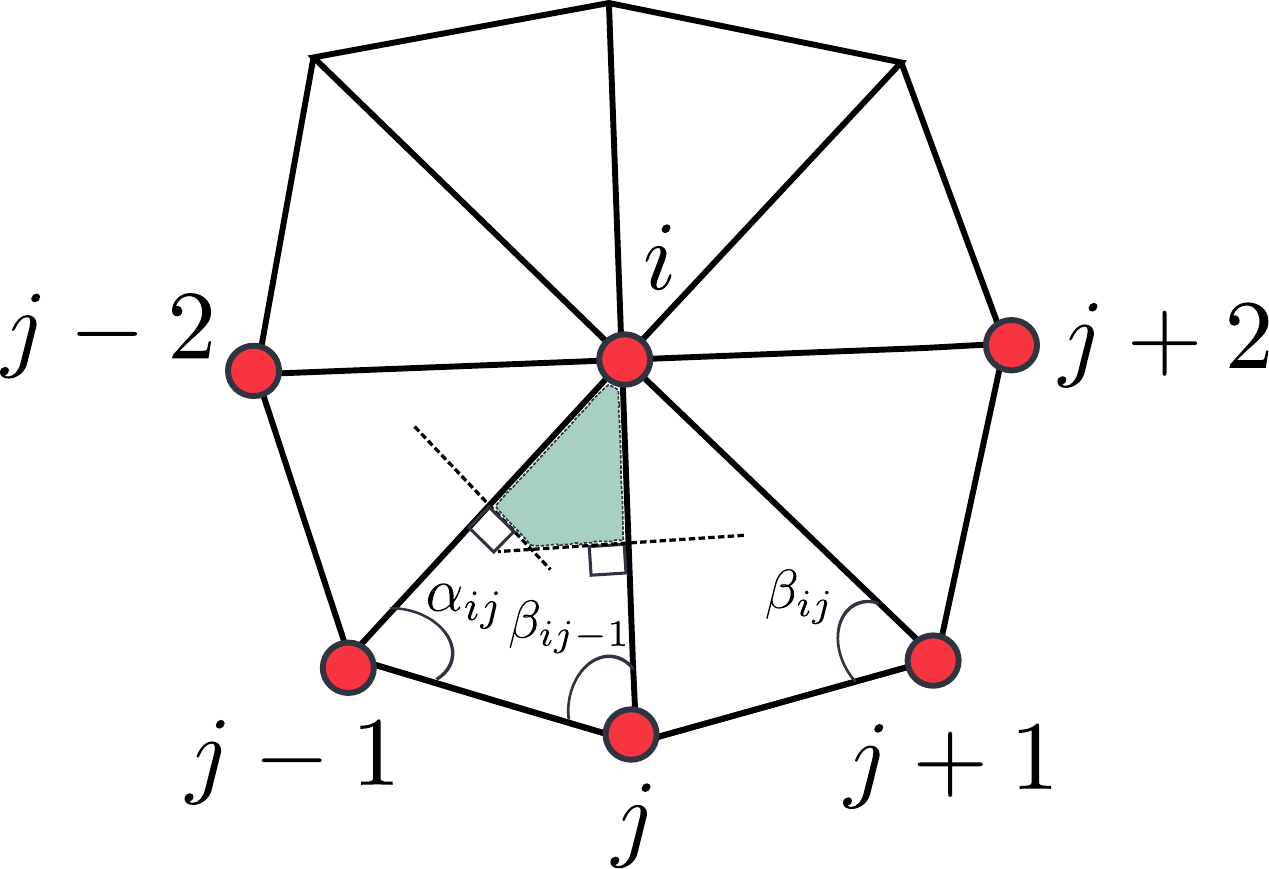}
  \caption{An example of triangulated mesh at the node $i$. 
  $\aij,\bij$ are the angles opposite to the bond $ij$.
  Shaded part is the Voronoi region of triangle T at the nodes $i,j-1,j$ -- it lies inside as T is non-obtuse.
  The nodes are sorted in counterclockwise direction.
  The image is adapted from Ref.~\cite{agrawal2022memc}.  
  }
  \label{fig:Voronoi}
\end{figure}
We use a Monte Carlo algorithm following Refs.~\cite{paulose2012fluctuating,
gompper2004triangulated} to study the elastic properties of shells. 
A version of our code, which performs thermal simulation of elastic shells, is
already available as an open source software~\cite{agrawal2022memc}.
Here we describe the algorithm in brief and  compare the results for 
elastic properties of shells in thermal equilibrium with those
obtained in Ref.~\cite{paulose2012fluctuating}.
How we adapt the same algorithm  to study shells out of equilibrium is
described in the main body of this paper. 
\subsection{Algorithm}
\label{app:algo}
We start with $N$ randomly chosen points on a sphere.
Then, we run a Monte Carlo simulation, with a repulsive
Lennard-Jones (LJ) potential, of these points moving on the surface
of the sphere.
Once the surface Monte Carlo (SMC) has reached thermal equilibrium,
we triangulate the points using the algorithm described in
Ref.~\cite{caroli2010robust}.
This is our initial configuration.
The distance between two neighboring nodes $i$ and $j$ is called $\lzij$.

We use Monte Carlo~\cite{baumgartner2013applications} simulations to update the
positions of the points ($\XXi$ at node $i$).
For a pressurized shell, the total energy
\begin{equation}
  E = \Es + \Eb + PV \/,
\label{eq:Etot}
\end{equation}
where the stretching and the bending 
contributions are, respectively, 
\begin{subequations}
  \begin{align}
    \Es &= \frac{1}{2}\Sum_i \frac{H}{2}\Sum_{j(i)}  \left(\XXij - \lzij\right)^2 \/, 
    \quad \text{where} 
    \label{eq:Es} \\
    \Xij &\equiv \lvert \XXi - \XXj \rvert \/\quad\text{and} \\
\Eb &= \frac{B}{2} \SUM_i \cAib \left(\LL_i - C\nhat\right)^2 \/.
  \label{eq:bendE}
  \end{align}
\end{subequations}
Here $P$ is the pressure difference between outside and inside the shell, 
and $V$ is the volume.
The Young's modulus of the membrane is given by
$Y=2H/\sqrt{3}$~\cite{itzykson1986proceedings}.
The bending modulus is $B$,
$\nhat$ is the outward normal to the surface, 
$C$ is its spontaneous curvature, 
and $\cAib$ is the area of Voronoi dual cell at the
node $i$~\cite{agrawal2022memc,
meyer2003discrete,hege2003visualization}.
The operator
\begin{equation}
  \LL_i = \frac{1}{\cAib}\SUM_{j(i)}
  \frac{1}{2}\left[\cot(\aij) + \cot(\bij)\right]\XXij\/,
\end{equation}
is the discrete Laplacian ~\cite{itzykson1986proceedings,
  hege2003visualization,meyer2003discrete} at the node $i$.
Here $\aij,\bij$ are the angles opposite to bond $\XXij$ as shown
in~\Fig{fig:Voronoi}. 
We compute $\cAib$ as follows~\cite{hege2003visualization,meyer2003discrete}.
Consider the triangle T in~\Fig{fig:Voronoi}, defined by the nodes $i,j,j-1$.
If T is non-obtuse, the area of shaded region in~\Fig{fig:Voronoi} is
\begin{equation}
  \cAjib = \frac{1}{8} \left[\Xij^2\cot(\aij) + \Xijm^2\cot(\bijm) \right]\/.
  \label{eq:cA}
\end{equation}
If T is an obtuse triangle, the shaded region
in~\Fig{fig:Voronoi} lies outside the triangle T, 
then if the angle at the vertex $i$ of T is obtuse
$\cAjib =\rm{area}(T)/2$ otherwise $\cAjib=\rm{area}(T)/4$.
Here  ${\rm area}(T) = (1/2)\lvert\XXij \times \XXijm\rvert$.
The area $\cAib$ is obtained by summing up the contributions from
all the triangles similar to $\cAjib$, in~\Fig{fig:Voronoi}, e.g.,
the contribution from the triangle T is the shaded area.

To compute the outward normal to the surface, $\nhat$, in~\Eq{eq:bendE},
we sort the points about node $i$ in a counterclockwise manner.
To sort the neighbors around any node $i$, we rotate the coordinate system
such that, the $z$ axis passes through the point $i$ along the vector $\XXi$.
In this coordinate system we sort the neighbors by their azimuthal angle. 
Note that unlike Ref.~\cite{gompper2004triangulated} we do not
incorporate self-avoidance.

\subsection{Comparison with Paulose \etal~\cite{paulose2012fluctuating}}
We use the radius of the sphere $R = 1$ as our unit of
length and $\kB T = 1$ as our unit of energy.

\begin{figure}[h]
  \includegraphics[width=\columnwidth]{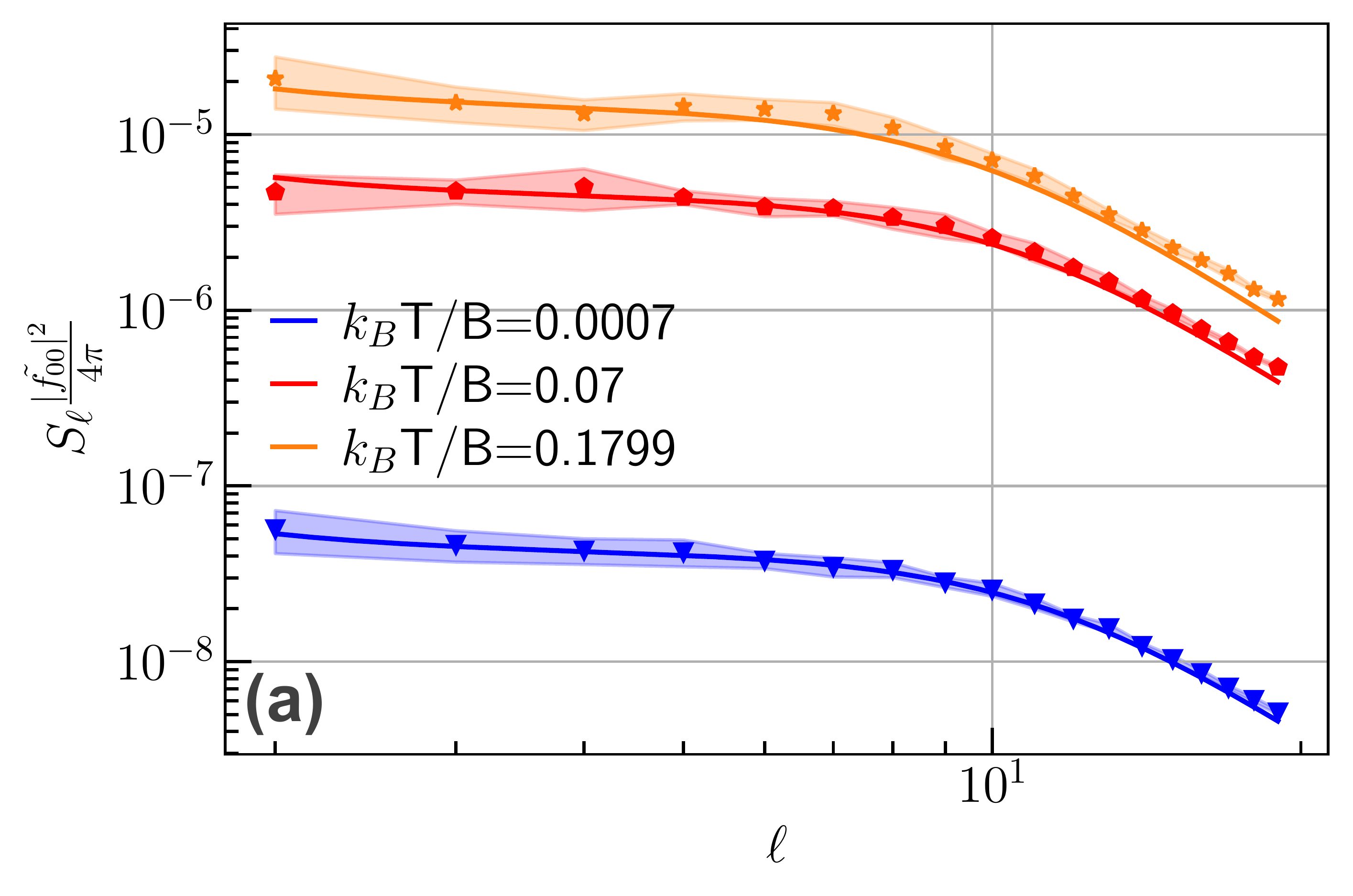}
  \includegraphics[width=\columnwidth]{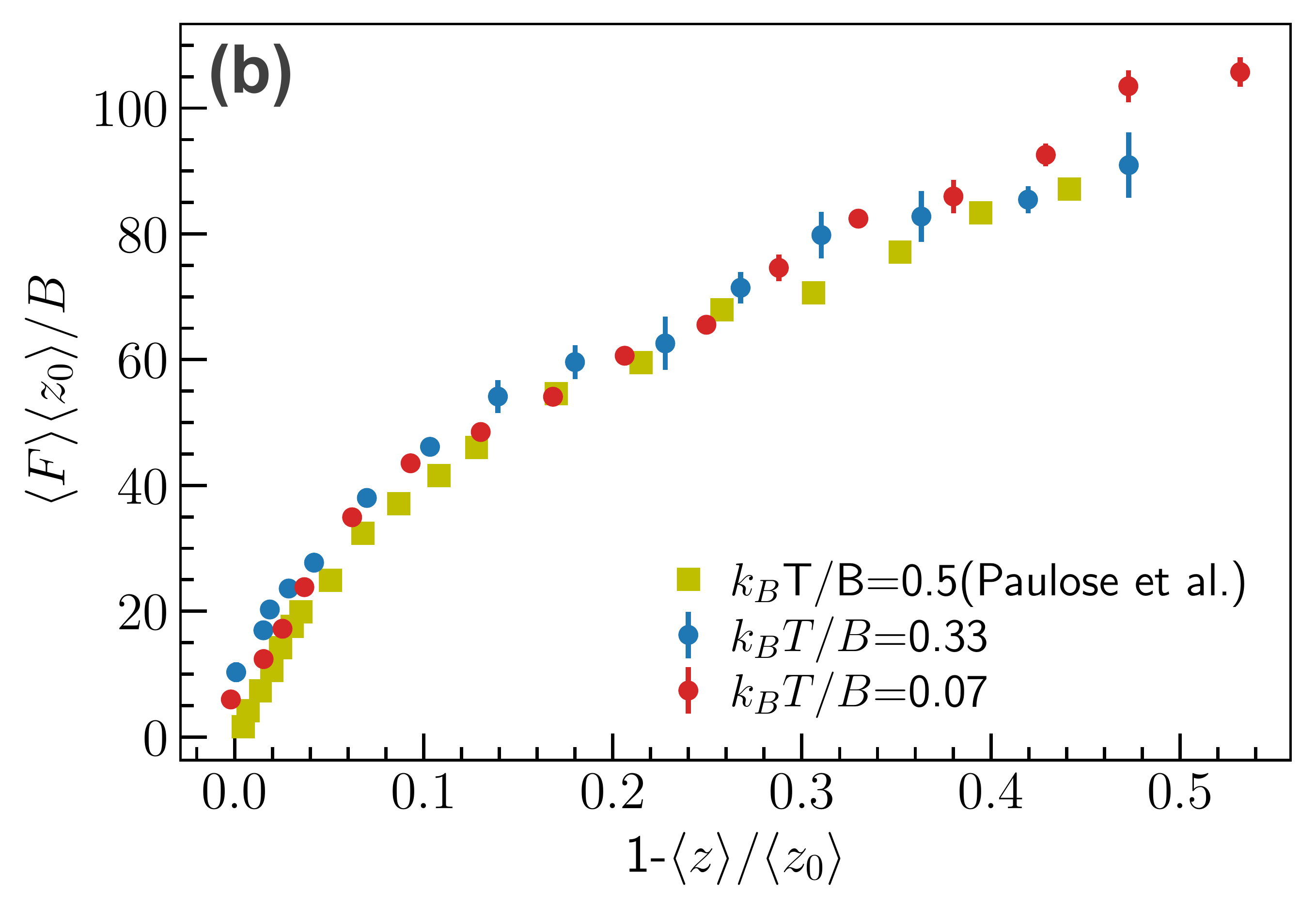}
  \caption{\textbf{Comparison with
      Ref~\citet{paulose2012fluctuating}:} 
      (a) The fluctuation spectrum $S(\ell)$ versus $\ell$
      for different values of bending modulus $B$. 
      The solid lines are theoretical predictions~\cite{paulose2012fluctuating}.
      The lines with symbols are from our simulations.
      The shaded area shows the error, see section~(\ref{sec:error}).
     (b) The force--deformation curve from our simulations compared with
     Ref.~\cite{paulose2012fluctuating} (yellow squares).
     We use  $N = 20292$ and $\FvK = 4616$.
  }
  \label{fig:vcode}
\end{figure}

We decompose the fluctuating height field, $f(\theta,\phi)$ in
spherical harmonics, $\Ylm(\theta,\phi)$:
\begin{subequations}
  \begin{align}
    f(\theta,\phi) &= \sum_{\ell,m}\flm\Ylm(\theta,\phi)\/, \quad
    \text{and define} \\
    S(\ell) &\equiv \frac{4\pi}{(2\ell+1)\lvert\ft_{00}\rvert^2}
    \sum_{m=-\ell}^{\ell}\lvert\flm\rvert^2 \/.
  \end{align}
  \label{eq:spec}
\end{subequations}
In \Fig{fig:vcode} we plot the spectra $S(\ell)$ for
number of nodes $N = 20252$,
the F\"oppl--von-Karman number $\FvK \equiv YR^2/B = 4616$ and 
for three values of the bending rigidity, 
$\kB T/B = 7\times10^{-4}$ (blue), $0.07$ (red) and $0.18$
(orange).
The continuous line shows the theoretical prediction by
Ref.~\cite{paulose2012fluctuating}
the shaded region is the error obtained by using BAGGing~(\ref{sec:error})
-- we find reasonable agreement with the theory.
Next, following Ref.~\cite{paulose2012fluctuating} we measure the  mechanical
response of a thermal shell by deforming it equally at the two poles
with two point-like indentations. 
This is implemented by two harmonic springs that 
are attached to the north and the south poles of the shell. 
In \subfig{fig:vcode}{B} we compare the force--deformation curve from our
simulations with those obtained in Ref.~\cite{paulose2012fluctuating}.
Again we obtain quite reasonable agreement.
A detailed comparison of parameters of our simulations with those
of Ref.~\cite{paulose2012fluctuating} is given in Table~\ref{tab:parameters}.
\begin{table}
\centering
\begin{tabular}{ | c | c | c | }
\hline
\textbf{Parameters} & \textbf{our simulation} & Ref.~\cite{paulose2012fluctuating}\\
\hline
Number of grid points, $N$ & $5120$ to $20252$ & $5530$ to $41816$  \\ 
\hline
Activity, $A$ & $-4$ to  $4$ & $0$ \\
\hline
$P/\Pnot$, & $-4$ to  $1$ & $0.2$ to $1$ \\
\hline
Elasto thermal number, $\ET$ & $1.65\times10^{-2}$ to $68$  & $10^{-6}$ to $10^2$ \\
\hline
\FVK, $\Fv$ & $4616$ & $650$ to $35000$  \\ 
\hline
Total MC steps & $10^6-5\times10^7$ & $1.25\times10^8$\\
\hline
\end{tabular}
\caption{Comparison between our simulations and those in
  Ref.~\cite{paulose2012fluctuating}.}
\label{tab:parameters}
\end{table}
\subsection{Comparison between Glauber and Metropolis algorithm}
\begin{figure}[h]
  \includegraphics[width=\columnwidth]{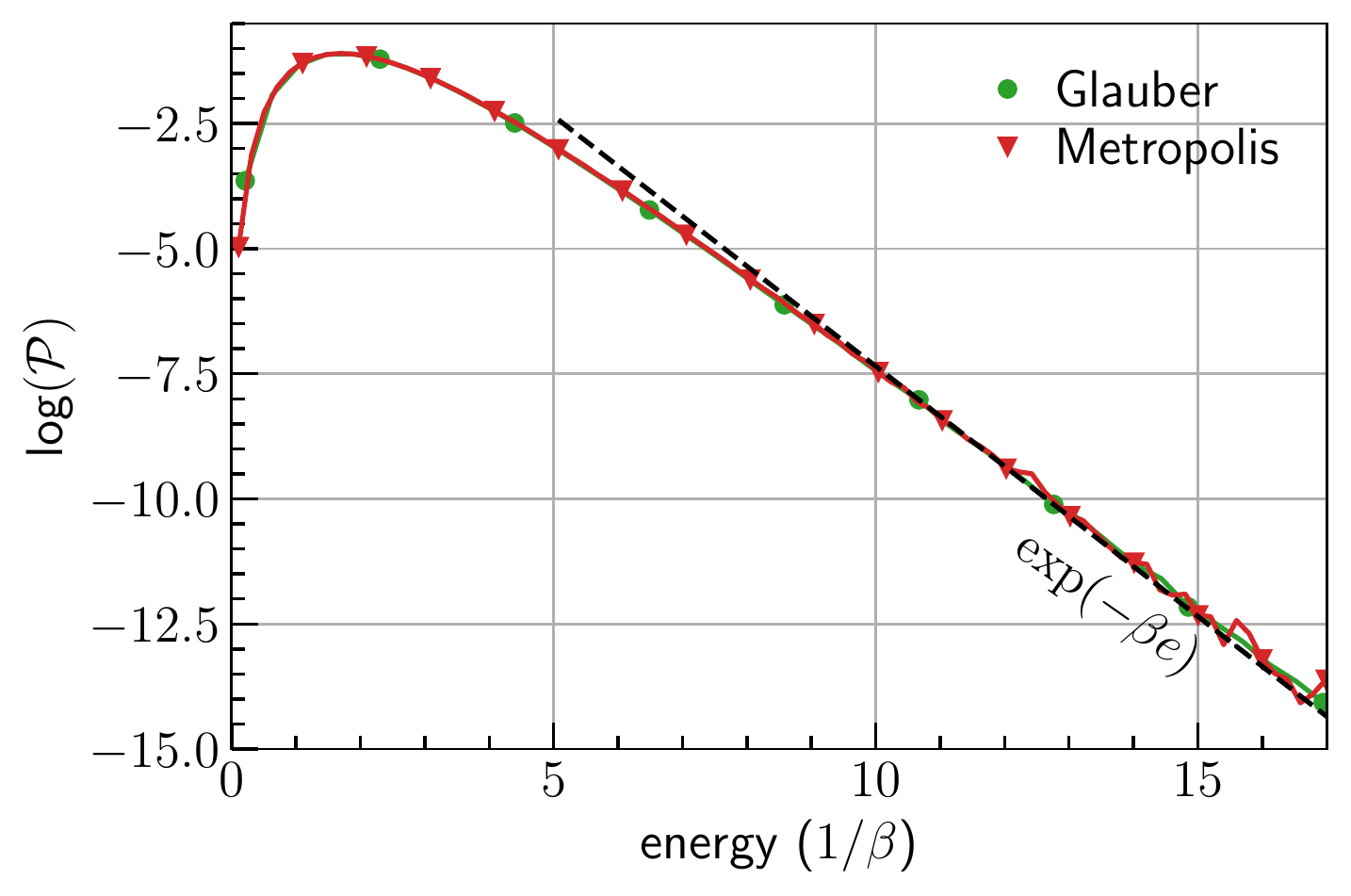}\\
  \caption{\textbf{Probability distribution function} of energy at each point 
      of the surface for Glauber and Metropolis algorithm for N=20252,
      and total MC steps = $10^6$.}
  \label{fig:gl_met}
\end{figure}

We run Monte Carlo simulations with
both Metropolis ($\WM$) and the Glauber ($\WG$),
rates:
\begin{subequations}
  \begin{align}
  \WM &= {\rm min}\left[1,\exp\left(-\frac{E}{\kB T}\right)\right]\/\text{and}
\label{eq:WM1}\\
  \WG &= \frac{1}{2}\left[1-\tanh\left(\frac{E}{2\kB T}\right)\right]\/.
  \label{eq:WG1}
  \end{align}
  \label{eq:W1}
\end{subequations}
Here $\kB$ is the Boltzmann constant, $T$ is the temperature 
and $E$ is the difference in energy between the two states.
In~\Fig{fig:gl_met}, we compare the probability distribution function (PDF) of
the energy (in thermal equilibrium) of individual nodes 
obtained by the Glauber and the Metropolis rules.
As expected, the PDF shows the same Boltzmann distribution
in both cases. 

\begin{figure}
\centering
    \includegraphics[width=\linewidth]{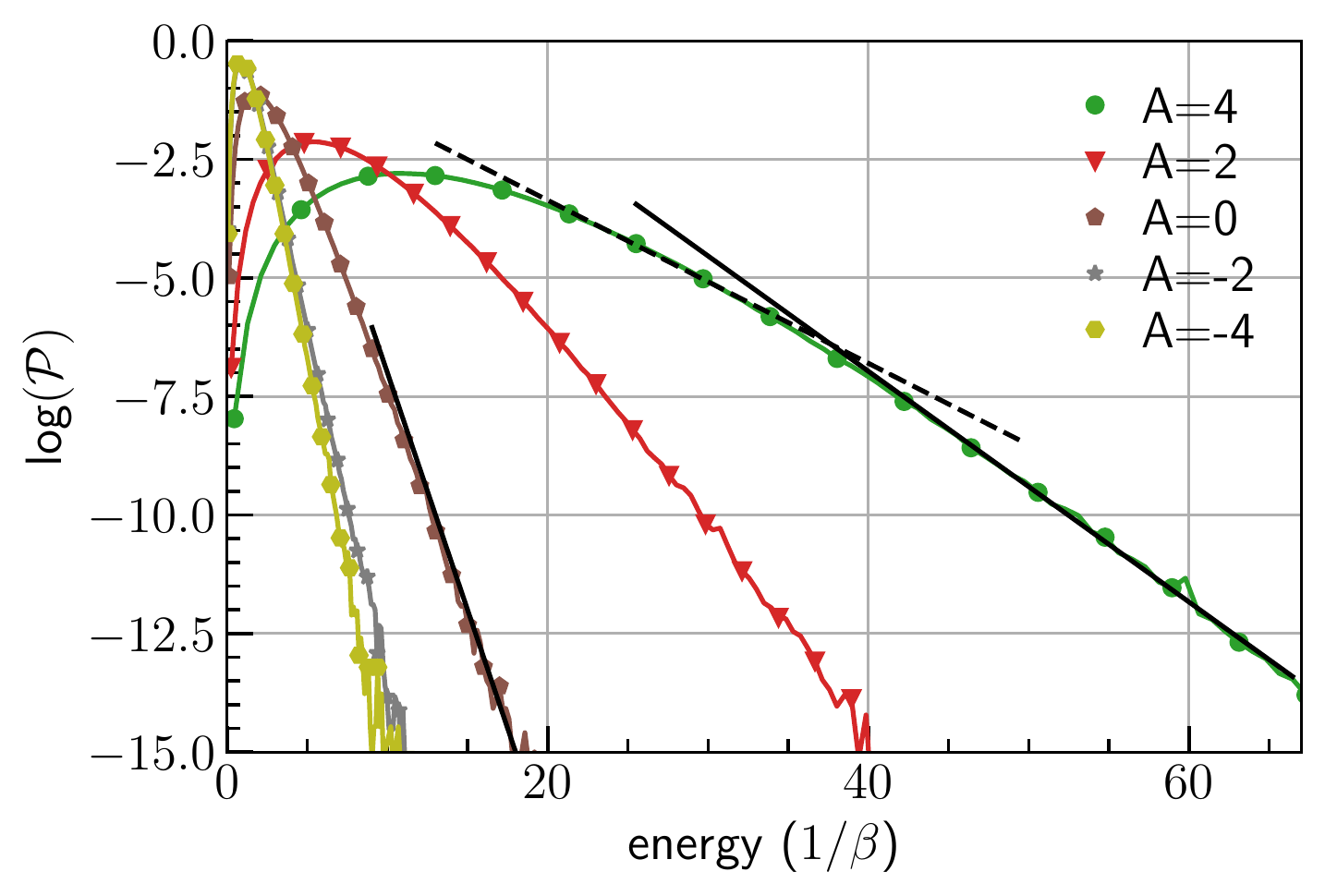}
    \caption{\label{fig:epoint}\textbf{Probability distribution
        function} of energies of every node in the stationary state for activity
      $A=-4, -2, 0, 2, 4$.
      The distribution fits an exponential distribution for the thermal (A=0)
      case. However for the active case (e.g. A=4), the distribution does not
      fit single exponential curve.}
\end{figure}
\section{Monte Carlo simulations out of equilibrium}
To drive the membrane out of equilibrium, following
Ref.~\cite{kumar2020nonequilibrium},
we replace $E$ in \eq{eq:W1} by $E + \Delta E$ where $\Delta E$ is a constant.
This guarantees that detailed balance is broken and the
amount by which it is broken is $\Delta E$.
If $\Delta E$ is positive (negative) the probability of acceptance
of large fluctuations is decreased (increased).
Thus we define a dimensionless quantity
$A = - \Delta E/(\kB T)$ such that simulations with
positive $A$, \textit{active} simulations, have higher fluctuations than
equilibrium ones whereas for negative $A$, \textit{sedate} simulations,
the fluctuations are less than the equilibrium ones.
A complete list of parameters are shown in caption of table~(\ref{tab:buck}). 

In \Fig{fig:epoint} we plot the probability distribution function (PDF) of
energy of individual nodes.
For the equilibrium case the tail of the PDF can be 
fitted with $\mP(E) \sim \exp-(E/\kB T)$,
where E is the energy of a node.
But for the out-of-equilibrium cases there is no one exponential that
can be used to fit.
\begin{figure}
\centering
    \includegraphics[width=0.95\linewidth]{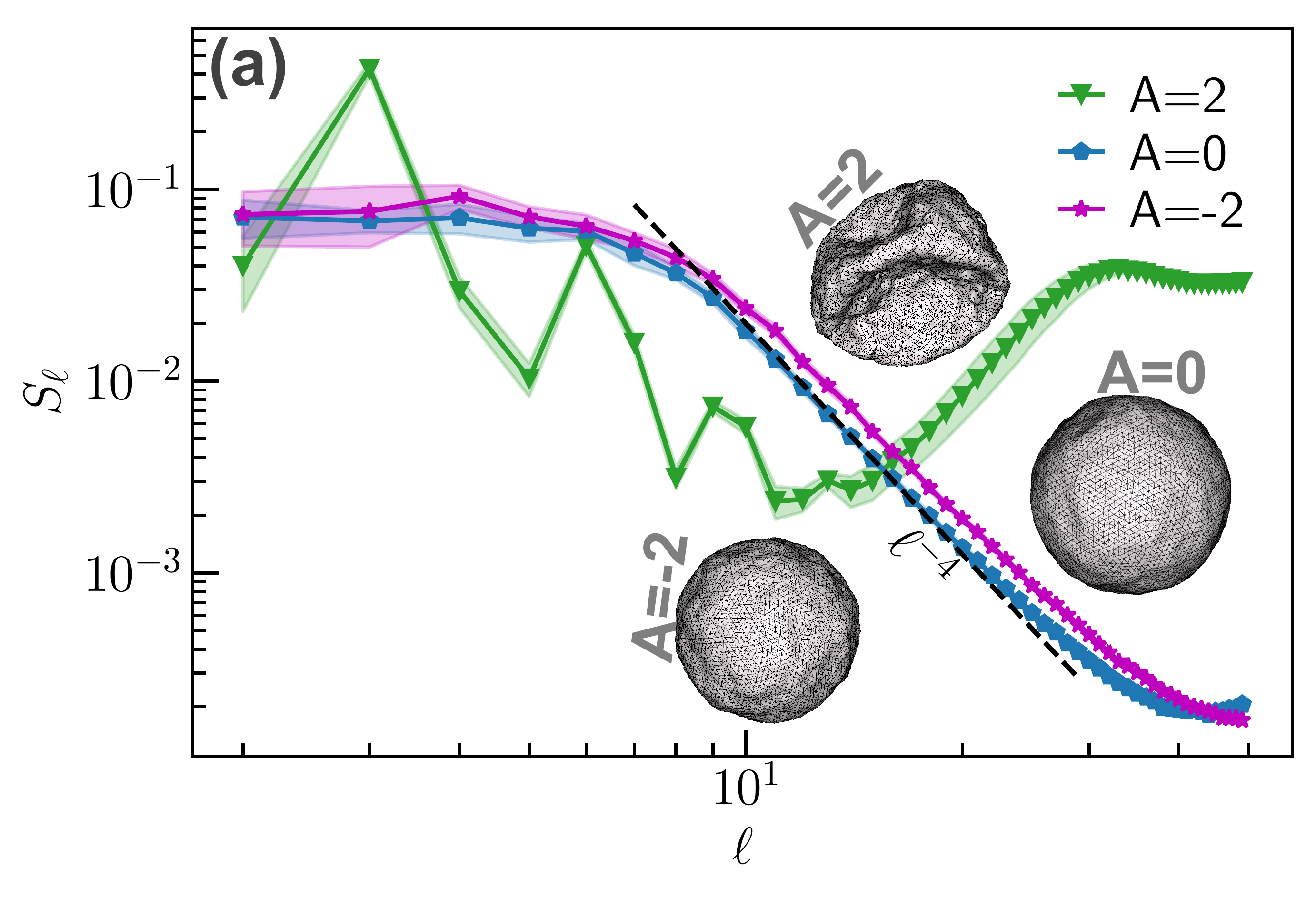} 
    \includegraphics[width=0.95\linewidth]{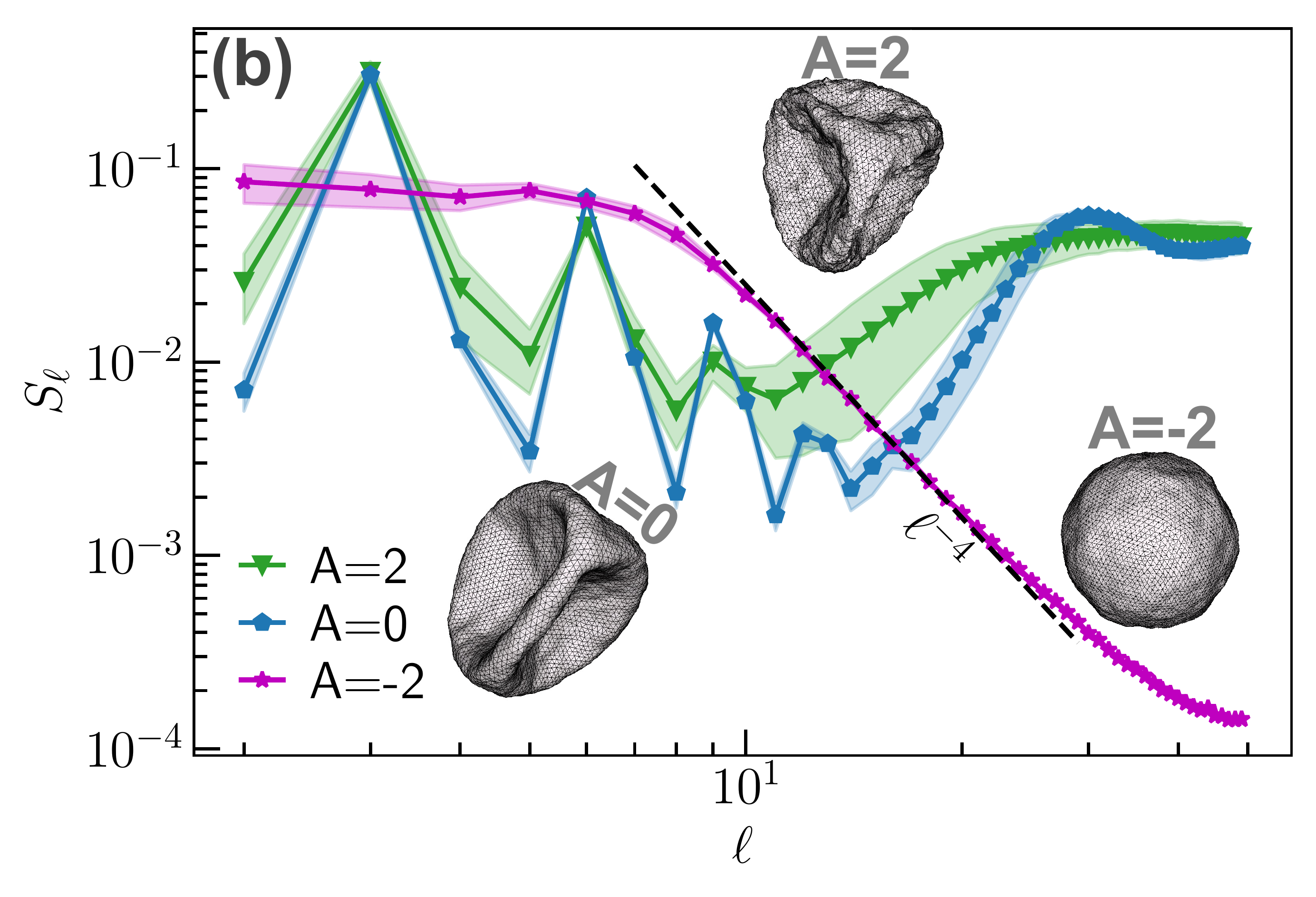}
    \caption{\label{fig:spectra}{\bf Spectra:} The spectra, $S(\ell)$,
      \eq{eq:spec}, $\ell\geq 2$ and (a) $P = 0.3\Pnot$ and (b) $P=0.36\Pnot$ for
      thermal (blue) active (green, $A=2$) and sedate~(magenta, $A=-2$).
      The solid lines are the mean of spectra  and
      shaded regions are the error.
      Appearance of a peak at small $\ell$ signifies buckling.
      } 
\end{figure}
The fluctuation spectrum $S(\ell)$ comparing unbuckled and buckled
simulations are shown in \Fig{fig:spectra}.
The unbuckled ones, even when out of equilibrium, shows
$S(\ell) \sim \ell^{-4}$ for a range of scales.
The buckled ones show distinctive peak at small values of $\ell$. 

To obtain the phase diagram we use thirteen values of elasto--thermal number,
for each of which we use seven values of activity.
For a fixed choice of elasto--thermal number and activity we start our
simulations with an initial condition where the shell is 
a perfect sphere. 
Then we choose a fixed value of external pressure and run our simulations
till we reach a stationary state, which for zero activity is the
equilibrium state. Whether the shell is buckled or not is decided
by three checks: (a) significant decrease of volume (b) a peak at
small $\ell$ for $S(\ell)$ (c) visual inspection.
If the shell in not buckled we choose a higher external pressure and
start our simulations again from the same initial condition. 
The buckling pressure, $\Pc$ obtained for a set of parameters is given
in table~(\ref{tab:buck}).
This way we mark out the phase boundary in the pressure--elasto--thermal
number plane for different activities and in the pressure--activity plane for
different elasto--thermal numbers, see Fig. 3 in main text.
In the Fig. 3(a) of the main text, we also plot the phase boundary, obtained
through a RG calculation in Ref.~\cite{kovsmrlj2017statistical},
which agrees reasonably well with our numerical results for zero activity. 
Note that for large enough values of $\ET$ and $A$ we reach a part
of the phase diagram where the shell is unstable at zero
external pressure and can be made stable only with positive internal
pressure. This part of the phase diagram is not shown,
although the relevant data are included in table~(\ref{tab:buck}).
Note that at small $\ET$ for the sedate case it is possible to
have the shell remain unbuckled even for pressure higher $\Pnot$,
i.e., the shell is stabilized.

\begin{table}
\begin{tabular}{|c|c|c|c|c|c|c|c|}
\hline
\backslashbox{\bET}{\bA} & 4 & 2 & 1 & 0 & -1 & -2 & -4\\\hline
$1.65\times10^{-2}$ & 0.70& 0.70& --& 0.70& 0.70& 0.80& 1.10\\\hline
$3.32\times10^{-2}$ & 0.60& 0.60& --& 0.70& 0.70& 0.80& 1.10\\\hline
$6.63\times10^{-2}$ & 0.60& 0.60& 0.60& 0.70& 0.70& 0.80& 1.00\\\hline
$1.33\times10^{-1}$ & 0.60& 0.60& 0.60& 0.70& 0.70& 0.80& 1.00\\\hline
$2.65\times10^{-1}$ & 0.50& 0.60& 0.60& 0.70& 0.70& 0.80& 1.00\\\hline
$5.31\times10^{-1}$ & 0.50& 0.50& 0.60& -- & 0.70& 0.80& 1.00\\\hline
$1.06$ &0.40& 0.40& 0.50& --& 0.60& 0.80& 1.00\\\hline
$2.12$ &0.40& 0.40& 0.47& 0.50& 0.60& 0.70& 1.00 \\\hline
$4.53$ &0.30& 0.30& --& 0.50& 0.50& 0.60& 0.80 \\\hline
$7.99$ &\underline{0.30}& \underline{0.30}& 0.30& 0.36& 0.40& 0.40& 0.50 \\\hline
$16.99$& -2.8& -- & --& \underline{0.30}& 0.40&0.40& -- \\\hline
$33.97$& $<$-4 & -2.5& --& 0.30& 0.30& 0.40& 0.50 \\\hline
$67.94$& $<$-4 & $<$-4 & -0.10& \underline{0.20}& 0.30& 0.30& 0.50 \\\hline
\end{tabular}
\caption{\label{tab:buck}{\bf Buckling pressure($\Pc/\Pnot$):}
  for different values of elasto--thermal number (\ET) and activity ($A$)
  obtained from the simulation with \FvK=4616,
  $N=5120$, $R=1$ and $\kB T=1$,
  where $N$ is the number of grid points, $R$ is radius of the shell,
  $\kB$ is the Boltzmann constant and $T$ is the temperature.
  We repeated some of the simulations for $N=20252$ and obtained the
  same buckling pressure. To reach a stationary state, typically,
  we need to run for $10^6-5\times10^7$ Monte Carlo steps,
  where in one Monte Carlo step, we perform $2N$ Monte Carlo iterations
  to attempt to update the position a random point.
  Underlined values have been reproduced with Glauber algorithm as well.
}
\end{table}
\subsection{Estimation of error}
\label{sec:error}
To compute the statistical error in a dataset $\Psi$, we
partition it into $M$ smaller datasets $\{\Psi_1, \ldots,\Psi_M\}$.
Then for each partition we calculate the statistical quantities,
e.g., mean, spectrum, etc.
The average over the partitions is the average value we quote and the
error is the standard deviation calculated across the partitions. 
We use $M=10$.
This technique is called BAGGing or Bootstrap AGGregation~\cite{mehta2019high}.

\begin{thebibliography}{63}%
\makeatletter
\providecommand \@ifxundefined [1]{%
 \@ifx{#1\undefined}
}%
\providecommand \@ifnum [1]{%
 \ifnum #1\expandafter \@firstoftwo
 \else \expandafter \@secondoftwo
 \fi
}%
\providecommand \@ifx [1]{%
 \ifx #1\expandafter \@firstoftwo
 \else \expandafter \@secondoftwo
 \fi
}%
\providecommand \natexlab [1]{#1}%
\providecommand \enquote  [1]{``#1''}%
\providecommand \bibnamefont  [1]{#1}%
\providecommand \bibfnamefont [1]{#1}%
\providecommand \citenamefont [1]{#1}%
\providecommand \href@noop [0]{\@secondoftwo}%
\providecommand \href [0]{\begingroup \@sanitize@url \@href}%
\providecommand \@href[1]{\@@startlink{#1}\@@href}%
\providecommand \@@href[1]{\endgroup#1\@@endlink}%
\providecommand \@sanitize@url [0]{\catcode `\\12\catcode `\$12\catcode
  `\&12\catcode `\#12\catcode `\^12\catcode `\_12\catcode `\%12\relax}%
\providecommand \@@startlink[1]{}%
\providecommand \@@endlink[0]{}%
\providecommand \url  [0]{\begingroup\@sanitize@url \@url }%
\providecommand \@url [1]{\endgroup\@href {#1}{\urlprefix }}%
\providecommand \urlprefix  [0]{URL }%
\providecommand \Eprint [0]{\href }%
\providecommand \doibase [0]{http://dx.doi.org/}%
\providecommand \selectlanguage [0]{\@gobble}%
\providecommand \bibinfo  [0]{\@secondoftwo}%
\providecommand \bibfield  [0]{\@secondoftwo}%
\providecommand \translation [1]{[#1]}%
\providecommand \BibitemOpen [0]{}%
\providecommand \bibitemStop [0]{}%
\providecommand \bibitemNoStop [0]{.\EOS\space}%
\providecommand \EOS [0]{\spacefactor3000\relax}%
\providecommand \BibitemShut  [1]{\csname bibitem#1\endcsname}%
\let\auto@bib@innerbib\@empty
\bibitem [{Note1()}]{Note1}%
  \BibitemOpen
  \bibinfo {note} {There are several geodesic domes with sizes ranging from 10
  to 200 meters.}\BibitemShut {Stop}%
\bibitem [{\citenamefont {Buenemann}\ and\ \citenamefont
  {Lenz}(2008)}]{buenemann2008elastic}%
  \BibitemOpen
  \bibfield  {author} {\bibinfo {author} {\bibfnamefont {Mathias}\ \bibnamefont
  {Buenemann}}\ and\ \bibinfo {author} {\bibfnamefont {Peter}\ \bibnamefont
  {Lenz}},\ }\bibfield  {title} {\enquote {\bibinfo {title} {Elastic properties
  and mechanical stability of chiral and filled viral capsids},}\ }\href@noop
  {} {\bibfield  {journal} {\bibinfo  {journal} {Physical Review E}\ }\textbf
  {\bibinfo {volume} {78}},\ \bibinfo {pages} {051924} (\bibinfo {year}
  {2008})}\BibitemShut {NoStop}%
\bibitem [{\citenamefont {Michel}\ \emph {et~al.}(2006)\citenamefont {Michel},
  \citenamefont {Ivanovska}, \citenamefont {Gibbons}, \citenamefont {Klug},
  \citenamefont {Knobler}, \citenamefont {Wuite},\ and\ \citenamefont
  {Schmidt}}]{michel2006nanoindentation}%
  \BibitemOpen
  \bibfield  {author} {\bibinfo {author} {\bibfnamefont {JP}~\bibnamefont
  {Michel}}, \bibinfo {author} {\bibfnamefont {IL}~\bibnamefont {Ivanovska}},
  \bibinfo {author} {\bibfnamefont {MM}~\bibnamefont {Gibbons}}, \bibinfo
  {author} {\bibfnamefont {WS}~\bibnamefont {Klug}}, \bibinfo {author}
  {\bibfnamefont {CM}~\bibnamefont {Knobler}}, \bibinfo {author} {\bibfnamefont
  {GJL}\ \bibnamefont {Wuite}}, \ and\ \bibinfo {author} {\bibfnamefont
  {CF}~\bibnamefont {Schmidt}},\ }\bibfield  {title} {\enquote {\bibinfo
  {title} {Nanoindentation studies of full and empty viral capsids and the
  effects of capsid protein mutations on elasticity and strength},}\
  }\href@noop {} {\bibfield  {journal} {\bibinfo  {journal} {Proceedings of the
  National Academy of Sciences}\ }\textbf {\bibinfo {volume} {103}},\ \bibinfo
  {pages} {6184--6189} (\bibinfo {year} {2006})}\BibitemShut {NoStop}%
\bibitem [{\citenamefont {Pegtel}\ and\ \citenamefont
  {Gould}(2019)}]{pegtel2019exosomes}%
  \BibitemOpen
  \bibfield  {author} {\bibinfo {author} {\bibfnamefont {D~Michiel}\
  \bibnamefont {Pegtel}}\ and\ \bibinfo {author} {\bibfnamefont {Stephen~J}\
  \bibnamefont {Gould}},\ }\bibfield  {title} {\enquote {\bibinfo {title}
  {Exosomes},}\ }\href@noop {} {\bibfield  {journal} {\bibinfo  {journal}
  {Annual review of biochemistry}\ }\textbf {\bibinfo {volume} {88}},\ \bibinfo
  {pages} {487--514} (\bibinfo {year} {2019})}\BibitemShut {NoStop}%
\bibitem [{\citenamefont {Cavallaro}\ \emph {et~al.}(2019)\citenamefont
  {Cavallaro}, \citenamefont {Horak}, \citenamefont {H{\aa}{\aa}g},
  \citenamefont {Gupta}, \citenamefont {Stiller}, \citenamefont {Sahu},
  \citenamefont {Gorgens}, \citenamefont {Gatty}, \citenamefont {Viktorsson},
  \citenamefont {El~Andaloussi} \emph {et~al.}}]{cavallaro2019label}%
  \BibitemOpen
  \bibfield  {author} {\bibinfo {author} {\bibfnamefont {Sara}\ \bibnamefont
  {Cavallaro}}, \bibinfo {author} {\bibfnamefont {Josef}\ \bibnamefont
  {Horak}}, \bibinfo {author} {\bibfnamefont {Petra}\ \bibnamefont
  {H{\aa}{\aa}g}}, \bibinfo {author} {\bibfnamefont {Dhanu}\ \bibnamefont
  {Gupta}}, \bibinfo {author} {\bibfnamefont {Christiane}\ \bibnamefont
  {Stiller}}, \bibinfo {author} {\bibfnamefont {Siddharth~S}\ \bibnamefont
  {Sahu}}, \bibinfo {author} {\bibfnamefont {Andre}\ \bibnamefont {Gorgens}},
  \bibinfo {author} {\bibfnamefont {Hithesh~K}\ \bibnamefont {Gatty}}, \bibinfo
  {author} {\bibfnamefont {Kristina}\ \bibnamefont {Viktorsson}}, \bibinfo
  {author} {\bibfnamefont {Samir}\ \bibnamefont {El~Andaloussi}},  \emph
  {et~al.},\ }\bibfield  {title} {\enquote {\bibinfo {title} {Label-free
  surface protein profiling of extracellular vesicles by an electrokinetic
  sensor},}\ }\href@noop {} {\bibfield  {journal} {\bibinfo  {journal} {ACS
  sensors}\ }\textbf {\bibinfo {volume} {4}},\ \bibinfo {pages} {1399--1408}
  (\bibinfo {year} {2019})}\BibitemShut {NoStop}%
\bibitem [{\citenamefont {Love}(2011)}]{Love2013treatise}%
  \BibitemOpen
  \bibfield  {author} {\bibinfo {author} {\bibfnamefont {Augustus
  Edward~Hough}\ \bibnamefont {Love}},\ }\href@noop {} {\emph {\bibinfo {title}
  {A treatise on the mathematical theory of elasticity}}}\ (\bibinfo
  {publisher} {Dover Publications; 4th edition},\ \bibinfo {year}
  {2011})\BibitemShut {NoStop}%
\bibitem [{\citenamefont {Landau}\ and\ \citenamefont
  {Lifshitz}(1970)}]{LLelast}%
  \BibitemOpen
  \bibfield  {author} {\bibinfo {author} {\bibfnamefont {LD}~\bibnamefont
  {Landau}}\ and\ \bibinfo {author} {\bibfnamefont {EM}~\bibnamefont
  {Lifshitz}},\ }\href@noop {} {\emph {\bibinfo {title} {Theory of
  Elasticity}}},\ \bibinfo {series} {Course of Theoretical Physics},
  Vol.~\bibinfo {volume} {7}\ (\bibinfo  {publisher} {Pergamon Press Ltd.},\
  \bibinfo {address} {Oxford, England},\ \bibinfo {year} {1970})\BibitemShut
  {NoStop}%
\bibitem [{\citenamefont {Koiter}(1963)}]{koiter1963progress}%
  \BibitemOpen
  \bibfield  {author} {\bibinfo {author} {\bibfnamefont {WT}~\bibnamefont
  {Koiter}},\ }\href@noop {} {\enquote {\bibinfo {title} {Progress in applied
  mechanics (the prager anniversary volume)},}\ } (\bibinfo {year}
  {1963})\BibitemShut {NoStop}%
\bibitem [{\citenamefont {WT}(1976)}]{koiter1976buckling}%
  \BibitemOpen
  \bibfield  {author} {\bibinfo {author} {\bibfnamefont {Koiter}\ \bibnamefont
  {WT}},\ }\bibfield  {title} {\enquote {\bibinfo {title} {Current trends in
  theory of buckling},}\ }in\ \href@noop {} {\emph {\bibinfo {booktitle}
  {Buckling of structures: Symposium Cambridge/USA}}},\ \bibinfo {editor}
  {edited by\ \bibinfo {editor} {\bibfnamefont {Bernard}\ \bibnamefont
  {Budiansky}}}\ (\bibinfo  {publisher} {Springer-Verlag},\ \bibinfo {address}
  {Berlin, Heidelberg, New York},\ \bibinfo {year} {1976})\ pp.\ \bibinfo
  {pages} {1--16}\BibitemShut {NoStop}%
\bibitem [{\citenamefont {Pogorelov}(1988)}]{pogorelov1988bendings}%
  \BibitemOpen
  \bibfield  {author} {\bibinfo {author} {\bibfnamefont {Alekse~Vasilevich}\
  \bibnamefont {Pogorelov}},\ }\href@noop {} {\emph {\bibinfo {title} {Bendings
  of surfaces and stability of shells}}},\ Vol.~\bibinfo {volume} {72}\
  (\bibinfo  {publisher} {American Mathematical Soc.},\ \bibinfo {year}
  {1988})\BibitemShut {NoStop}%
\bibitem [{\citenamefont {Hutchinson}(2016)}]{hutchinson2016buckling}%
  \BibitemOpen
  \bibfield  {author} {\bibinfo {author} {\bibfnamefont {John~W}\ \bibnamefont
  {Hutchinson}},\ }\bibfield  {title} {\enquote {\bibinfo {title} {Buckling of
  spherical shells revisited},}\ }\href@noop {} {\bibfield  {journal} {\bibinfo
   {journal} {Proceedings of the Royal Society A: Mathematical, Physical and
  Engineering Sciences}\ }\textbf {\bibinfo {volume} {472}},\ \bibinfo {pages}
  {20160577} (\bibinfo {year} {2016})}\BibitemShut {NoStop}%
\bibitem [{\citenamefont {Vliegenthart}\ and\ \citenamefont
  {Gompper}(2006)}]{vliegenthart2006mechanical}%
  \BibitemOpen
  \bibfield  {author} {\bibinfo {author} {\bibfnamefont {Gerard~Adriaan}\
  \bibnamefont {Vliegenthart}}\ and\ \bibinfo {author} {\bibfnamefont
  {Gerhard}\ \bibnamefont {Gompper}},\ }\bibfield  {title} {\enquote {\bibinfo
  {title} {Mechanical deformation of spherical viruses with icosahedral
  symmetry},}\ }\href@noop {} {\bibfield  {journal} {\bibinfo  {journal}
  {Biophysical journal}\ }\textbf {\bibinfo {volume} {91}},\ \bibinfo {pages}
  {834--841} (\bibinfo {year} {2006})}\BibitemShut {NoStop}%
\bibitem [{\citenamefont {Gao}\ \emph {et~al.}(2001)\citenamefont {Gao},
  \citenamefont {Donath}, \citenamefont {Moya}, \citenamefont {Dudnik},\ and\
  \citenamefont {M{\"o}hwald}}]{gao2001elasticity}%
  \BibitemOpen
  \bibfield  {author} {\bibinfo {author} {\bibfnamefont {Ch}~\bibnamefont
  {Gao}}, \bibinfo {author} {\bibfnamefont {E}~\bibnamefont {Donath}}, \bibinfo
  {author} {\bibfnamefont {S}~\bibnamefont {Moya}}, \bibinfo {author}
  {\bibfnamefont {V}~\bibnamefont {Dudnik}}, \ and\ \bibinfo {author}
  {\bibfnamefont {H}~\bibnamefont {M{\"o}hwald}},\ }\bibfield  {title}
  {\enquote {\bibinfo {title} {Elasticity of hollow polyelectrolyte capsules
  prepared by the layer-by-layer technique},}\ }\href@noop {} {\bibfield
  {journal} {\bibinfo  {journal} {The European Physical Journal E}\ }\textbf
  {\bibinfo {volume} {5}},\ \bibinfo {pages} {21--27} (\bibinfo {year}
  {2001})}\BibitemShut {NoStop}%
\bibitem [{\citenamefont {Gordon}\ \emph {et~al.}(2004)\citenamefont {Gordon},
  \citenamefont {Chen}, \citenamefont {Hutchinson}, \citenamefont {Bausch},
  \citenamefont {Marquez},\ and\ \citenamefont {Weitz}}]{gordon2004self}%
  \BibitemOpen
  \bibfield  {author} {\bibinfo {author} {\bibfnamefont {Vernita~D}\
  \bibnamefont {Gordon}}, \bibinfo {author} {\bibfnamefont {Xi}~\bibnamefont
  {Chen}}, \bibinfo {author} {\bibfnamefont {John~W}\ \bibnamefont
  {Hutchinson}}, \bibinfo {author} {\bibfnamefont {Andreas~R}\ \bibnamefont
  {Bausch}}, \bibinfo {author} {\bibfnamefont {Manuel}\ \bibnamefont
  {Marquez}}, \ and\ \bibinfo {author} {\bibfnamefont {David~A}\ \bibnamefont
  {Weitz}},\ }\bibfield  {title} {\enquote {\bibinfo {title} {Self-assembled
  polymer membrane capsules inflated by osmotic pressure},}\ }\href@noop {}
  {\bibfield  {journal} {\bibinfo  {journal} {Journal of the American Chemical
  Society}\ }\textbf {\bibinfo {volume} {126}},\ \bibinfo {pages}
  {14117--14122} (\bibinfo {year} {2004})}\BibitemShut {NoStop}%
\bibitem [{\citenamefont {Vella}\ \emph {et~al.}(2012)\citenamefont {Vella},
  \citenamefont {Ajdari}, \citenamefont {Vaziri},\ and\ \citenamefont
  {Boudaoud}}]{vella2012indentation}%
  \BibitemOpen
  \bibfield  {author} {\bibinfo {author} {\bibfnamefont {Dominic}\ \bibnamefont
  {Vella}}, \bibinfo {author} {\bibfnamefont {Amin}\ \bibnamefont {Ajdari}},
  \bibinfo {author} {\bibfnamefont {Ashkan}\ \bibnamefont {Vaziri}}, \ and\
  \bibinfo {author} {\bibfnamefont {Arezki}\ \bibnamefont {Boudaoud}},\
  }\bibfield  {title} {\enquote {\bibinfo {title} {The indentation of
  pressurized elastic shells: from polymeric capsules to yeast cells},}\
  }\href@noop {} {\bibfield  {journal} {\bibinfo  {journal} {Journal of the
  Royal Society Interface}\ }\textbf {\bibinfo {volume} {9}},\ \bibinfo {pages}
  {448--455} (\bibinfo {year} {2012})}\BibitemShut {NoStop}%
\bibitem [{\citenamefont {Vorselen}\ \emph {et~al.}(2017)\citenamefont
  {Vorselen}, \citenamefont {MacKintosh}, \citenamefont {Roos},\ and\
  \citenamefont {Wuite}}]{vorselen2017competition}%
  \BibitemOpen
  \bibfield  {author} {\bibinfo {author} {\bibfnamefont {Daan}\ \bibnamefont
  {Vorselen}}, \bibinfo {author} {\bibfnamefont {Fred~C}\ \bibnamefont
  {MacKintosh}}, \bibinfo {author} {\bibfnamefont {Wouter~H}\ \bibnamefont
  {Roos}}, \ and\ \bibinfo {author} {\bibfnamefont {Gijs~JL}\ \bibnamefont
  {Wuite}},\ }\bibfield  {title} {\enquote {\bibinfo {title} {Competition
  between bending and internal pressure governs the mechanics of fluid
  nanovesicles},}\ }\href@noop {} {\bibfield  {journal} {\bibinfo  {journal}
  {Acs Nano}\ }\textbf {\bibinfo {volume} {11}},\ \bibinfo {pages} {2628--2636}
  (\bibinfo {year} {2017})}\BibitemShut {NoStop}%
\bibitem [{\citenamefont {Wu}\ \emph {et~al.}(2018)\citenamefont {Wu},
  \citenamefont {Aroush}, \citenamefont {Asnacios}, \citenamefont {Chen},
  \citenamefont {Dokukin}, \citenamefont {Doss}, \citenamefont {Durand-Smet},
  \citenamefont {Ekpenyong}, \citenamefont {Guck}, \citenamefont {Guz} \emph
  {et~al.}}]{wu2018comparison}%
  \BibitemOpen
  \bibfield  {author} {\bibinfo {author} {\bibfnamefont {Pei-Hsun}\
  \bibnamefont {Wu}}, \bibinfo {author} {\bibfnamefont {Dikla Raz-Ben}\
  \bibnamefont {Aroush}}, \bibinfo {author} {\bibfnamefont {Atef}\ \bibnamefont
  {Asnacios}}, \bibinfo {author} {\bibfnamefont {Wei-Chiang}\ \bibnamefont
  {Chen}}, \bibinfo {author} {\bibfnamefont {Maxim~E}\ \bibnamefont {Dokukin}},
  \bibinfo {author} {\bibfnamefont {Bryant~L}\ \bibnamefont {Doss}}, \bibinfo
  {author} {\bibfnamefont {Pauline}\ \bibnamefont {Durand-Smet}}, \bibinfo
  {author} {\bibfnamefont {Andrew}\ \bibnamefont {Ekpenyong}}, \bibinfo
  {author} {\bibfnamefont {Jochen}\ \bibnamefont {Guck}}, \bibinfo {author}
  {\bibfnamefont {Nataliia~V}\ \bibnamefont {Guz}},  \emph {et~al.},\
  }\bibfield  {title} {\enquote {\bibinfo {title} {A comparison of methods to
  assess cell mechanical properties},}\ }\href@noop {} {\bibfield  {journal}
  {\bibinfo  {journal} {Nature methods}\ }\textbf {\bibinfo {volume} {15}},\
  \bibinfo {pages} {491--498} (\bibinfo {year} {2018})}\BibitemShut {NoStop}%
\bibitem [{\citenamefont {Vorselen}\ \emph {et~al.}(2020)\citenamefont
  {Vorselen}, \citenamefont {Piontek}, \citenamefont {Roos},\ and\
  \citenamefont {Wuite}}]{vorselen2020mechanical}%
  \BibitemOpen
  \bibfield  {author} {\bibinfo {author} {\bibfnamefont {Daan}\ \bibnamefont
  {Vorselen}}, \bibinfo {author} {\bibfnamefont {Melissa~C}\ \bibnamefont
  {Piontek}}, \bibinfo {author} {\bibfnamefont {Wouter~H}\ \bibnamefont
  {Roos}}, \ and\ \bibinfo {author} {\bibfnamefont {Gijs~JL}\ \bibnamefont
  {Wuite}},\ }\bibfield  {title} {\enquote {\bibinfo {title} {Mechanical
  characterization of liposomes and extracellular vesicles, a protocol},}\
  }\href@noop {} {\bibfield  {journal} {\bibinfo  {journal} {Frontiers in
  molecular biosciences}\ }\textbf {\bibinfo {volume} {7}},\ \bibinfo {pages}
  {139} (\bibinfo {year} {2020})}\BibitemShut {NoStop}%
\bibitem [{\citenamefont {Cavallaro}\ \emph {et~al.}(2021)\citenamefont
  {Cavallaro}, \citenamefont {Pevere}, \citenamefont {Stridfeldt},
  \citenamefont {G{\"o}rgens}, \citenamefont {Paba}, \citenamefont {Sahu},
  \citenamefont {Mamand}, \citenamefont {Gupta}, \citenamefont {El~Andaloussi},
  \citenamefont {Linnros} \emph {et~al.}}]{cavallaro2021multiparametric}%
  \BibitemOpen
  \bibfield  {author} {\bibinfo {author} {\bibfnamefont {Sara}\ \bibnamefont
  {Cavallaro}}, \bibinfo {author} {\bibfnamefont {Federico}\ \bibnamefont
  {Pevere}}, \bibinfo {author} {\bibfnamefont {Fredrik}\ \bibnamefont
  {Stridfeldt}}, \bibinfo {author} {\bibfnamefont {Andr{\'e}}\ \bibnamefont
  {G{\"o}rgens}}, \bibinfo {author} {\bibfnamefont {Carolina}\ \bibnamefont
  {Paba}}, \bibinfo {author} {\bibfnamefont {Siddharth~S}\ \bibnamefont
  {Sahu}}, \bibinfo {author} {\bibfnamefont {Doste~R}\ \bibnamefont {Mamand}},
  \bibinfo {author} {\bibfnamefont {Dhanu}\ \bibnamefont {Gupta}}, \bibinfo
  {author} {\bibfnamefont {Samir}\ \bibnamefont {El~Andaloussi}}, \bibinfo
  {author} {\bibfnamefont {Jan}\ \bibnamefont {Linnros}},  \emph {et~al.},\
  }\bibfield  {title} {\enquote {\bibinfo {title} {Multiparametric profiling of
  single nanoscale extracellular vesicles by combined atomic force and
  fluorescence microscopy: correlation and heterogeneity in their molecular and
  biophysical features},}\ }\href@noop {} {\bibfield  {journal} {\bibinfo
  {journal} {Small}\ }\textbf {\bibinfo {volume} {17}},\ \bibinfo {pages}
  {2008155} (\bibinfo {year} {2021})}\BibitemShut {NoStop}%
\bibitem [{\citenamefont {Zoldesi}\ \emph {et~al.}(2008)\citenamefont
  {Zoldesi}, \citenamefont {Ivanovska}, \citenamefont {Quilliet}, \citenamefont
  {Wuite},\ and\ \citenamefont {Imhof}}]{zoldesi2008elastic}%
  \BibitemOpen
  \bibfield  {author} {\bibinfo {author} {\bibfnamefont {CI}~\bibnamefont
  {Zoldesi}}, \bibinfo {author} {\bibfnamefont {IL}~\bibnamefont {Ivanovska}},
  \bibinfo {author} {\bibfnamefont {C}~\bibnamefont {Quilliet}}, \bibinfo
  {author} {\bibfnamefont {GJL}\ \bibnamefont {Wuite}}, \ and\ \bibinfo
  {author} {\bibfnamefont {A}~\bibnamefont {Imhof}},\ }\bibfield  {title}
  {\enquote {\bibinfo {title} {Elastic properties of hollow colloidal
  particles},}\ }\href@noop {} {\bibfield  {journal} {\bibinfo  {journal}
  {Physical Review E}\ }\textbf {\bibinfo {volume} {78}},\ \bibinfo {pages}
  {051401} (\bibinfo {year} {2008})}\BibitemShut {NoStop}%
\bibitem [{\citenamefont {Jackson}\ \emph {et~al.}(2022)\citenamefont
  {Jackson}, \citenamefont {Romeo}, \citenamefont {Mietke}, \citenamefont
  {Burns}, \citenamefont {Totz}, \citenamefont {Martin}, \citenamefont
  {Dunkel},\ and\ \citenamefont {Alsous}}]{jackson2022dynamics}%
  \BibitemOpen
  \bibfield  {author} {\bibinfo {author} {\bibfnamefont {Jonathan~A}\
  \bibnamefont {Jackson}}, \bibinfo {author} {\bibfnamefont {Nicolas}\
  \bibnamefont {Romeo}}, \bibinfo {author} {\bibfnamefont {Alexander}\
  \bibnamefont {Mietke}}, \bibinfo {author} {\bibfnamefont {Keaton~J}\
  \bibnamefont {Burns}}, \bibinfo {author} {\bibfnamefont {Jan~F}\ \bibnamefont
  {Totz}}, \bibinfo {author} {\bibfnamefont {Adam~C}\ \bibnamefont {Martin}},
  \bibinfo {author} {\bibfnamefont {J{\"o}rn}\ \bibnamefont {Dunkel}}, \ and\
  \bibinfo {author} {\bibfnamefont {Jasmin~Imran}\ \bibnamefont {Alsous}},\
  }\bibfield  {title} {\enquote {\bibinfo {title} {Dynamics, scaling behavior,
  and control of nuclear wrinkling},}\ }\href@noop {} {\bibfield  {journal}
  {\bibinfo  {journal} {arXiv preprint arXiv:2210.11581}\ } (\bibinfo {year}
  {2022})}\BibitemShut {NoStop}%
\bibitem [{\citenamefont {HW}\ \emph {et~al.}(2002)\citenamefont {HW},
  \citenamefont {Wortis},\ and\ \citenamefont
  {Mukhopadhyay}}]{hw2002stomatocyte}%
  \BibitemOpen
  \bibfield  {author} {\bibinfo {author} {\bibfnamefont {Gerald~Lim}\
  \bibnamefont {HW}}, \bibinfo {author} {\bibfnamefont {Michael}\ \bibnamefont
  {Wortis}}, \ and\ \bibinfo {author} {\bibfnamefont {Ranjan}\ \bibnamefont
  {Mukhopadhyay}},\ }\bibfield  {title} {\enquote {\bibinfo {title}
  {Stomatocyte--discocyte--echinocyte sequence of the human red blood cell:
  Evidence for the bilayer--couple hypothesis from membrane mechanics},}\
  }\href@noop {} {\bibfield  {journal} {\bibinfo  {journal} {Proceedings of the
  National Academy of Sciences}\ }\textbf {\bibinfo {volume} {99}},\ \bibinfo
  {pages} {16766--16769} (\bibinfo {year} {2002})}\BibitemShut {NoStop}%
\bibitem [{\citenamefont {Fedosov}\ \emph
  {et~al.}(2010{\natexlab{a}})\citenamefont {Fedosov}, \citenamefont
  {Caswell},\ and\ \citenamefont {Karniadakis}}]{fedosov2010multiscale}%
  \BibitemOpen
  \bibfield  {author} {\bibinfo {author} {\bibfnamefont {Dmitry~A}\
  \bibnamefont {Fedosov}}, \bibinfo {author} {\bibfnamefont {Bruce}\
  \bibnamefont {Caswell}}, \ and\ \bibinfo {author} {\bibfnamefont {George~Em}\
  \bibnamefont {Karniadakis}},\ }\bibfield  {title} {\enquote {\bibinfo {title}
  {A multiscale red blood cell model with accurate mechanics, rheology, and
  dynamics},}\ }\href@noop {} {\bibfield  {journal} {\bibinfo  {journal}
  {Biophysical journal}\ }\textbf {\bibinfo {volume} {98}},\ \bibinfo {pages}
  {2215--2225} (\bibinfo {year} {2010}{\natexlab{a}})}\BibitemShut {NoStop}%
\bibitem [{\citenamefont {Fedosov}\ \emph
  {et~al.}(2010{\natexlab{b}})\citenamefont {Fedosov}, \citenamefont
  {Caswell},\ and\ \citenamefont {Karniadakis}}]{fedosov2010systematic}%
  \BibitemOpen
  \bibfield  {author} {\bibinfo {author} {\bibfnamefont {Dmitry~A}\
  \bibnamefont {Fedosov}}, \bibinfo {author} {\bibfnamefont {Bruce}\
  \bibnamefont {Caswell}}, \ and\ \bibinfo {author} {\bibfnamefont {George~Em}\
  \bibnamefont {Karniadakis}},\ }\bibfield  {title} {\enquote {\bibinfo {title}
  {Systematic coarse-graining of spectrin-level red blood cell models},}\
  }\href@noop {} {\bibfield  {journal} {\bibinfo  {journal} {Computer Methods
  in Applied Mechanics and Engineering}\ }\textbf {\bibinfo {volume} {199}},\
  \bibinfo {pages} {1937--1948} (\bibinfo {year}
  {2010}{\natexlab{b}})}\BibitemShut {NoStop}%
\bibitem [{\citenamefont {Baumg{\"a}rtner}\ \emph {et~al.}(2013)\citenamefont
  {Baumg{\"a}rtner}, \citenamefont {Binder}, \citenamefont {Hansen},
  \citenamefont {Kalos}, \citenamefont {Kehr}, \citenamefont {Landau},
  \citenamefont {Levesque}, \citenamefont {M{\"u}ller-Krumbhaar}, \citenamefont
  {Rebbi}, \citenamefont {Saito} \emph {et~al.}}]{baumgartner2013applications}%
  \BibitemOpen
  \bibfield  {author} {\bibinfo {author} {\bibfnamefont {Dmitry~A}\
  \bibnamefont {Fedosov}}, \bibinfo {author} {\bibfnamefont {Bruce}\
  \bibnamefont {Caswell}}, \ and\ \bibinfo {author} {\bibfnamefont {George~Em}\
  \bibnamefont {Karniadakis}},\ }\bibfield  {title} {\enquote {\bibinfo {title}
  {Systematic coarse-graining of spectrin-level red blood cell models},}\
  }\href@noop {} {\bibfield  {journal} {\bibinfo  {journal} {Computer Methods
  in Applied Mechanics and Engineering}\ }\textbf {\bibinfo {volume} {199}},\
  \bibinfo {pages} {1937--1948} (\bibinfo {year}
  {2010}{\natexlab{b}})}\BibitemShut {NoStop}%
\bibitem [{\citenamefont {Fedosov}(2010)}]{Fedosov2010multi}%
  \BibitemOpen
  \bibfield  {author} {\bibinfo {author} {\bibfnamefont {Dmitry~A}\
  \bibnamefont {Fedosov}},\ }\emph {\bibinfo {title} {Multiscale modeling of
  blood flow and soft matter}},\ \href@noop {} {Ph.D. thesis},\ \bibinfo
  {school} {Division of Applied Mathematics at Brown University} (\bibinfo
  {year} {2010})\BibitemShut {NoStop}%
\bibitem [{\citenamefont {Fedosov}\ \emph {et~al.}(2011)\citenamefont
  {Fedosov}, \citenamefont {Pan}, \citenamefont {Caswell}, \citenamefont
  {Gompper},\ and\ \citenamefont {Karniadakis}}]{fedosov2011predicting}%
  \BibitemOpen
  \bibfield  {author} {\bibinfo {author} {\bibfnamefont {Dmitry~A}\
  \bibnamefont {Fedosov}}, \bibinfo {author} {\bibfnamefont {Wenxiao}\
  \bibnamefont {Pan}}, \bibinfo {author} {\bibfnamefont {Bruce}\ \bibnamefont
  {Caswell}}, \bibinfo {author} {\bibfnamefont {Gerhard}\ \bibnamefont
  {Gompper}}, \ and\ \bibinfo {author} {\bibfnamefont {George~E}\ \bibnamefont
  {Karniadakis}},\ }\bibfield  {title} {\enquote {\bibinfo {title} {Predicting
  human blood viscosity in silico},}\ }\href@noop {} {\bibfield  {journal}
  {\bibinfo  {journal} {Proceedings of the National Academy of Sciences}\
  }\textbf {\bibinfo {volume} {108}},\ \bibinfo {pages} {11772--11777}
  (\bibinfo {year} {2011})}\BibitemShut {NoStop}%
\bibitem [{\citenamefont {Freund}(2014)}]{freund2014numerical}%
  \BibitemOpen
  \bibfield  {author} {\bibinfo {author} {\bibfnamefont {Jonathan~B}\
  \bibnamefont {Freund}},\ }\bibfield  {title} {\enquote {\bibinfo {title}
  {Numerical simulation of flowing blood cells},}\ }\href@noop {} {\bibfield
  {journal} {\bibinfo  {journal} {Annual review of fluid mechanics}\ }\textbf
  {\bibinfo {volume} {46}},\ \bibinfo {pages} {67--95} (\bibinfo {year}
  {2014})}\BibitemShut {NoStop}%
\bibitem [{\citenamefont {Zhu}\ \emph {et~al.}(2014)\citenamefont {Zhu},
  \citenamefont {Rorai}, \citenamefont {Mitra},\ and\ \citenamefont
  {Brandt}}]{zhu2014microfluidic}%
  \BibitemOpen
  \bibfield  {author} {\bibinfo {author} {\bibfnamefont {Lailai}\ \bibnamefont
  {Zhu}}, \bibinfo {author} {\bibfnamefont {Cecilia}\ \bibnamefont {Rorai}},
  \bibinfo {author} {\bibfnamefont {Dhrubaditya}\ \bibnamefont {Mitra}}, \ and\
  \bibinfo {author} {\bibfnamefont {Luca}\ \bibnamefont {Brandt}},\ }\bibfield
  {title} {\enquote {\bibinfo {title} {A microfluidic device to sort capsules
  by deformability: a numerical study},}\ }\href@noop {} {\bibfield  {journal}
  {\bibinfo  {journal} {Soft Matter}\ }\textbf {\bibinfo {volume} {10}},\
  \bibinfo {pages} {7705--7711} (\bibinfo {year} {2014})}\BibitemShut {NoStop}%
\bibitem [{\citenamefont {Paulose}\ \emph {et~al.}(2012)\citenamefont
  {Paulose}, \citenamefont {Vliegenthart}, \citenamefont {Gompper},\ and\
  \citenamefont {Nelson}}]{paulose2012fluctuating}%
  \BibitemOpen
  \bibfield  {author} {\bibinfo {author} {\bibfnamefont {Jayson}\ \bibnamefont
  {Paulose}}, \bibinfo {author} {\bibfnamefont {Gerard~A}\ \bibnamefont
  {Vliegenthart}}, \bibinfo {author} {\bibfnamefont {Gerhard}\ \bibnamefont
  {Gompper}}, \ and\ \bibinfo {author} {\bibfnamefont {David~R}\ \bibnamefont
  {Nelson}},\ }\bibfield  {title} {\enquote {\bibinfo {title} {Fluctuating
  shells under pressure},}\ }\href@noop {} {\bibfield  {journal} {\bibinfo
  {journal} {Proceedings of the National Academy of Sciences}\ }\textbf
  {\bibinfo {volume} {109}},\ \bibinfo {pages} {19551--19556} (\bibinfo {year}
  {2012})}\BibitemShut {NoStop}%
\bibitem [{\citenamefont {Ko{\v{s}}mrlj}\ and\ \citenamefont
  {Nelson}(2017)}]{kovsmrlj2017statistical}%
  \BibitemOpen
  \bibfield  {author} {\bibinfo {author} {\bibfnamefont {Andrej}\ \bibnamefont
  {Ko{\v{s}}mrlj}}\ and\ \bibinfo {author} {\bibfnamefont {David~R}\
  \bibnamefont {Nelson}},\ }\bibfield  {title} {\enquote {\bibinfo {title}
  {Statistical mechanics of thin spherical shells},}\ }\href@noop {} {\bibfield
   {journal} {\bibinfo  {journal} {Physical Review X}\ }\textbf {\bibinfo
  {volume} {7}},\ \bibinfo {pages} {011002} (\bibinfo {year}
  {2017})}\BibitemShut {NoStop}%
\bibitem [{\citenamefont {Schr\"odinger}(2012)}]{schrodinger2012life}%
  \BibitemOpen
  \bibfield  {author} {\bibinfo {author} {\bibfnamefont {Erwin}\ \bibnamefont
  {Schr\"odinger}},\ }\href@noop {} {\emph {\bibinfo {title} {What is life?:
  With mind and matter and autobiographical sketches}}}\ (\bibinfo  {publisher}
  {Cambridge university press},\ \bibinfo {year} {2012})\BibitemShut {NoStop}%
\bibitem [{\citenamefont {Gnesotto}\ \emph {et~al.}(2018)\citenamefont
  {Gnesotto}, \citenamefont {Mura}, \citenamefont {Gladrow},\ and\
  \citenamefont {Broedersz}}]{gnesotto2018broken}%
  \BibitemOpen
  \bibfield  {author} {\bibinfo {author} {\bibfnamefont {F~S}\ \bibnamefont
  {Gnesotto}}, \bibinfo {author} {\bibfnamefont {F}~\bibnamefont {Mura}},
  \bibinfo {author} {\bibfnamefont {J}~\bibnamefont {Gladrow}}, \ and\ \bibinfo
  {author} {\bibfnamefont {C~P}\ \bibnamefont {Broedersz}},\ }\bibfield
  {title} {\enquote {\bibinfo {title} {Broken detailed balance and
  non-equilibrium dynamics in living systems: a review},}\ }\href@noop {}
  {\bibfield  {journal} {\bibinfo  {journal} {Reports on Progress in Physics}\
  }\textbf {\bibinfo {volume} {81}},\ \bibinfo {pages} {066601} (\bibinfo
  {year} {2018})}\BibitemShut {NoStop}%
\bibitem [{\citenamefont {Marchetti}\ \emph {et~al.}(2013)\citenamefont
  {Marchetti}, \citenamefont {Joanny}, \citenamefont {Ramaswamy}, \citenamefont
  {Liverpool}, \citenamefont {Prost}, \citenamefont {Rao},\ and\ \citenamefont
  {Simha}}]{marchetti2013hydrodynamics}%
  \BibitemOpen
  \bibfield  {author} {\bibinfo {author} {\bibfnamefont {M~Cristina}\
  \bibnamefont {Marchetti}}, \bibinfo {author} {\bibfnamefont
  {Jean-Fran{\c{c}}ois}\ \bibnamefont {Joanny}}, \bibinfo {author}
  {\bibfnamefont {Sriram}\ \bibnamefont {Ramaswamy}}, \bibinfo {author}
  {\bibfnamefont {Tanniemola~B}\ \bibnamefont {Liverpool}}, \bibinfo {author}
  {\bibfnamefont {Jacques}\ \bibnamefont {Prost}}, \bibinfo {author}
  {\bibfnamefont {Madan}\ \bibnamefont {Rao}}, \ and\ \bibinfo {author}
  {\bibfnamefont {R~Aditi}\ \bibnamefont {Simha}},\ }\bibfield  {title}
  {\enquote {\bibinfo {title} {Hydrodynamics of soft active matter},}\
  }\href@noop {} {\bibfield  {journal} {\bibinfo  {journal} {Reviews of modern
  physics}\ }\textbf {\bibinfo {volume} {85}},\ \bibinfo {pages} {1143}
  (\bibinfo {year} {2013})}\BibitemShut {NoStop}%
\bibitem [{\citenamefont {Ramaswamy}(2010)}]{ramaswamy2010mechanics}%
  \BibitemOpen
  \bibfield  {author} {\bibinfo {author} {\bibfnamefont {Sriram}\ \bibnamefont
  {Ramaswamy}},\ }\bibfield  {title} {\enquote {\bibinfo {title} {The mechanics
  and statistics of active matter},}\ }\href@noop {} {\bibfield  {journal}
  {\bibinfo  {journal} {The Annual Review of Condensed Matter Physics is}\
  }\textbf {\bibinfo {volume} {1}},\ \bibinfo {pages} {323--45} (\bibinfo
  {year} {2010})}\BibitemShut {NoStop}%
\bibitem [{\citenamefont {Peng}\ \emph {et~al.}(2013)\citenamefont {Peng},
  \citenamefont {Li}, \citenamefont {Pivkin}, \citenamefont {Dao},
  \citenamefont {Karniadakis},\ and\ \citenamefont {Suresh}}]{peng2013lipid}%
  \BibitemOpen
  \bibfield  {author} {\bibinfo {author} {\bibfnamefont {Zhangli}\ \bibnamefont
  {Peng}}, \bibinfo {author} {\bibfnamefont {Xuejin}\ \bibnamefont {Li}},
  \bibinfo {author} {\bibfnamefont {Igor~V}\ \bibnamefont {Pivkin}}, \bibinfo
  {author} {\bibfnamefont {Ming}\ \bibnamefont {Dao}}, \bibinfo {author}
  {\bibfnamefont {George~E}\ \bibnamefont {Karniadakis}}, \ and\ \bibinfo
  {author} {\bibfnamefont {Subra}\ \bibnamefont {Suresh}},\ }\bibfield  {title}
  {\enquote {\bibinfo {title} {Lipid bilayer and cytoskeletal interactions in a
  red blood cell},}\ }\href@noop {} {\bibfield  {journal} {\bibinfo  {journal}
  {Proceedings of the National Academy of Sciences}\ }\textbf {\bibinfo
  {volume} {110}},\ \bibinfo {pages} {13356--13361} (\bibinfo {year}
  {2013})}\BibitemShut {NoStop}%
\bibitem [{\citenamefont {Turlier}\ \emph {et~al.}(2016)\citenamefont
  {Turlier}, \citenamefont {Fedosov}, \citenamefont {Audoly}, \citenamefont
  {Auth}, \citenamefont {Gov}, \citenamefont {Sykes}, \citenamefont {Joanny},
  \citenamefont {Gompper},\ and\ \citenamefont
  {Betz}}]{turlier2016equilibrium}%
  \BibitemOpen
  \bibfield  {author} {\bibinfo {author} {\bibfnamefont {Herv{\'e}}\
  \bibnamefont {Turlier}}, \bibinfo {author} {\bibfnamefont {Dmitry~A}\
  \bibnamefont {Fedosov}}, \bibinfo {author} {\bibfnamefont {Basile}\
  \bibnamefont {Audoly}}, \bibinfo {author} {\bibfnamefont {Thorsten}\
  \bibnamefont {Auth}}, \bibinfo {author} {\bibfnamefont {Nir~S}\ \bibnamefont
  {Gov}}, \bibinfo {author} {\bibfnamefont {C{\'e}cile}\ \bibnamefont {Sykes}},
  \bibinfo {author} {\bibfnamefont {J-F}\ \bibnamefont {Joanny}}, \bibinfo
  {author} {\bibfnamefont {Gerhard}\ \bibnamefont {Gompper}}, \ and\ \bibinfo
  {author} {\bibfnamefont {Timo}\ \bibnamefont {Betz}},\ }\bibfield  {title}
  {\enquote {\bibinfo {title} {Equilibrium physics breakdown reveals the active
  nature of red blood cell flickering},}\ }\href@noop {} {\bibfield  {journal}
  {\bibinfo  {journal} {Nature physics}\ }\textbf {\bibinfo {volume} {12}},\
  \bibinfo {pages} {513--519} (\bibinfo {year} {2016})}\BibitemShut {NoStop}%
\bibitem [{\citenamefont {Biswas}\ \emph {et~al.}(2017)\citenamefont {Biswas},
  \citenamefont {Alex},\ and\ \citenamefont {Sinha}}]{biswas2017mapping}%
  \BibitemOpen
  \bibfield  {author} {\bibinfo {author} {\bibfnamefont {Arikta}\ \bibnamefont
  {Biswas}}, \bibinfo {author} {\bibfnamefont {Amal}\ \bibnamefont {Alex}}, \
  and\ \bibinfo {author} {\bibfnamefont {Bidisha}\ \bibnamefont {Sinha}},\
  }\bibfield  {title} {\enquote {\bibinfo {title} {Mapping cell membrane
  fluctuations reveals their active regulation and transient
  heterogeneities},}\ }\href@noop {} {\bibfield  {journal} {\bibinfo  {journal}
  {Biophysical journal}\ }\textbf {\bibinfo {volume} {113}},\ \bibinfo {pages}
  {1768--1781} (\bibinfo {year} {2017})}\BibitemShut {NoStop}%
\bibitem [{\citenamefont {Turlier}\ and\ \citenamefont
  {Betz}(2019)}]{turlier2019unveiling}%
  \BibitemOpen
  \bibfield  {author} {\bibinfo {author} {\bibfnamefont {Herv{\'e}}\
  \bibnamefont {Turlier}}\ and\ \bibinfo {author} {\bibfnamefont {Timo}\
  \bibnamefont {Betz}},\ }\bibfield  {title} {\enquote {\bibinfo {title}
  {Unveiling the active nature of living-membrane fluctuations and
  mechanics},}\ }\href@noop {} {\bibfield  {journal} {\bibinfo  {journal}
  {Annual Review of Condensed Matter Physics}\ }\textbf {\bibinfo {volume}
  {10}},\ \bibinfo {pages} {213--232} (\bibinfo {year} {2019})}\BibitemShut
  {NoStop}%
\bibitem [{\citenamefont {Manikandan}\ \emph {et~al.}(2022)\citenamefont
  {Manikandan}, \citenamefont {Ghosh}, \citenamefont {Mandal}, \citenamefont
  {Biswas}, \citenamefont {Sinha},\ and\ \citenamefont
  {Mitra}}]{manikandan2022estimate}%
  \BibitemOpen
  \bibfield  {author} {\bibinfo {author} {\bibfnamefont {Sreekanth~K}\
  \bibnamefont {Manikandan}}, \bibinfo {author} {\bibfnamefont {Tanmoy}\
  \bibnamefont {Ghosh}}, \bibinfo {author} {\bibfnamefont {Tithi}\ \bibnamefont
  {Mandal}}, \bibinfo {author} {\bibfnamefont {Arikta}\ \bibnamefont {Biswas}},
  \bibinfo {author} {\bibfnamefont {Bidisha}\ \bibnamefont {Sinha}}, \ and\
  \bibinfo {author} {\bibfnamefont {Dhrubaditya}\ \bibnamefont {Mitra}},\
  }\bibfield  {title} {\enquote {\bibinfo {title} {Estimate of entropy
  generation rate can spatiotemporally resolve the active nature of cell
  flickering},}\ }\href@noop {} {\bibfield  {journal} {\bibinfo  {journal}
  {arXiv preprint arXiv:2205.12849}\ } (\bibinfo {year} {2022})}\BibitemShut
  {NoStop}%
\bibitem [{\citenamefont {Girard}\ \emph {et~al.}(2005)\citenamefont {Girard},
  \citenamefont {Prost},\ and\ \citenamefont {Bassereau}}]{girard2005passive}%
  \BibitemOpen
  \bibfield  {author} {\bibinfo {author} {\bibfnamefont {P}~\bibnamefont
  {Girard}}, \bibinfo {author} {\bibfnamefont {J}~\bibnamefont {Prost}}, \ and\
  \bibinfo {author} {\bibfnamefont {P}~\bibnamefont {Bassereau}},\ }\bibfield
  {title} {\enquote {\bibinfo {title} {Passive or active fluctuations in
  membranes containing proteins},}\ }\href@noop {} {\bibfield  {journal}
  {\bibinfo  {journal} {Physical review letters}\ }\textbf {\bibinfo {volume}
  {94}},\ \bibinfo {pages} {088102} (\bibinfo {year} {2005})}\BibitemShut
  {NoStop}%
\bibitem [{\citenamefont {Manneville}\ \emph {et~al.}(2001)\citenamefont
  {Manneville}, \citenamefont {Bassereau}, \citenamefont {Ramaswamy},\ and\
  \citenamefont {Prost}}]{manneville2001active}%
  \BibitemOpen
  \bibfield  {author} {\bibinfo {author} {\bibfnamefont {J-B}\ \bibnamefont
  {Manneville}}, \bibinfo {author} {\bibfnamefont {P}~\bibnamefont
  {Bassereau}}, \bibinfo {author} {\bibfnamefont {S}~\bibnamefont {Ramaswamy}},
  \ and\ \bibinfo {author} {\bibfnamefont {J}~\bibnamefont {Prost}},\
  }\bibfield  {title} {\enquote {\bibinfo {title} {Active membrane fluctuations
  studied by micropipet aspiration},}\ }\href@noop {} {\bibfield  {journal}
  {\bibinfo  {journal} {Physical Review E}\ }\textbf {\bibinfo {volume} {64}},\
  \bibinfo {pages} {021908} (\bibinfo {year} {2001})}\BibitemShut {NoStop}%
\bibitem [{\citenamefont {Yan}\ \emph {et~al.}(2021)\citenamefont {Yan},
  \citenamefont {Pezzulla}, \citenamefont {Cruveiller}, \citenamefont
  {Abbasi},\ and\ \citenamefont {Reis}}]{yan2021magneto}%
  \BibitemOpen
  \bibfield  {author} {\bibinfo {author} {\bibfnamefont {Dong}\ \bibnamefont
  {Yan}}, \bibinfo {author} {\bibfnamefont {Matteo}\ \bibnamefont {Pezzulla}},
  \bibinfo {author} {\bibfnamefont {Lilian}\ \bibnamefont {Cruveiller}},
  \bibinfo {author} {\bibfnamefont {Arefeh}\ \bibnamefont {Abbasi}}, \ and\
  \bibinfo {author} {\bibfnamefont {Pedro~M}\ \bibnamefont {Reis}},\ }\bibfield
   {title} {\enquote {\bibinfo {title} {Magneto-active elastic shells with
  tunable buckling strength},}\ }\href@noop {} {\bibfield  {journal} {\bibinfo
  {journal} {Nature communications}\ }\textbf {\bibinfo {volume} {12}},\
  \bibinfo {pages} {1--9} (\bibinfo {year} {2021})}\BibitemShut {NoStop}%
\bibitem [{\citenamefont {Kumar}\ and\ \citenamefont
  {Dasgupta}(2020)}]{kumar2020nonequilibrium}%
  \BibitemOpen
  \bibfield  {author} {\bibinfo {author} {\bibfnamefont {Manoj}\ \bibnamefont
  {Kumar}}\ and\ \bibinfo {author} {\bibfnamefont {Chandan}\ \bibnamefont
  {Dasgupta}},\ }\bibfield  {title} {\enquote {\bibinfo {title} {Nonequilibrium
  phase transition in an ising model without detailed balance},}\ }\href@noop
  {} {\bibfield  {journal} {\bibinfo  {journal} {Physical Review E}\ }\textbf
  {\bibinfo {volume} {102}},\ \bibinfo {pages} {052111} (\bibinfo {year}
  {2020})}\BibitemShut {NoStop}%
\bibitem [{\citenamefont {Gompper}\ and\ \citenamefont
  {Kroll}(2004)}]{gompper2004triangulated}%
  \BibitemOpen
  \bibfield  {author} {\bibinfo {author} {\bibfnamefont {G}~\bibnamefont
  {Gompper}}\ and\ \bibinfo {author} {\bibfnamefont {DM}~\bibnamefont
  {Kroll}},\ }\bibfield  {title} {\enquote {\bibinfo {title}
  {Triangulated-surface models of fluctuating membranes},}\ }in\ \href@noop {}
  {\emph {\bibinfo {booktitle} {Statistical mechanics of membranes and
  surfaces}}},\ \bibinfo {editor} {edited by\ \bibinfo {editor} {\bibfnamefont
  {David}\ \bibnamefont {Nelson}}, \bibinfo {editor} {\bibfnamefont {Tsvi}\
  \bibnamefont {Piran}}, \ and\ \bibinfo {editor} {\bibfnamefont {Steven}\
  \bibnamefont {Weinberg}}}\ (\bibinfo  {publisher} {World Scientific},\
  \bibinfo {year} {2004})\ pp.\ \bibinfo {pages} {359--426}\BibitemShut
  {NoStop}%
\bibitem [{\citenamefont {Mehta}\ \emph {et~al.}(2019)\citenamefont {Mehta},
  \citenamefont {Bukov}, \citenamefont {Wang}, \citenamefont {Day},
  \citenamefont {Richardson}, \citenamefont {Fisher},\ and\ \citenamefont
  {Schwab}}]{mehta2019high}%
  \BibitemOpen
  \bibfield  {author} {\bibinfo {author} {\bibfnamefont {Pankaj}\ \bibnamefont
  {Mehta}}, \bibinfo {author} {\bibfnamefont {Marin}\ \bibnamefont {Bukov}},
  \bibinfo {author} {\bibfnamefont {Ching-Hao}\ \bibnamefont {Wang}}, \bibinfo
  {author} {\bibfnamefont {Alexandre~GR}\ \bibnamefont {Day}}, \bibinfo
  {author} {\bibfnamefont {Clint}\ \bibnamefont {Richardson}}, \bibinfo
  {author} {\bibfnamefont {Charles~K}\ \bibnamefont {Fisher}}, \ and\ \bibinfo
  {author} {\bibfnamefont {David~J}\ \bibnamefont {Schwab}},\ }\bibfield
  {title} {\enquote {\bibinfo {title} {A high-bias, low-variance introduction
  to machine learning for physicists},}\ }\href@noop {} {\bibfield  {journal}
  {\bibinfo  {journal} {Physics reports}\ }\textbf {\bibinfo {volume} {810}},\
  \bibinfo {pages} {1--124} (\bibinfo {year} {2019})}\BibitemShut {NoStop}%
\bibitem [{\citenamefont {Itzykson}(1986)}]{itzykson1986proceedings}%
  \BibitemOpen
  \bibfield  {author} {\bibinfo {author} {\bibfnamefont {C}~\bibnamefont
  {Itzykson}},\ }in\ \href@noop {} {\emph {\bibinfo {booktitle} {Proceedings of
  the GIFT seminar, Jaca 85}}},\ \bibinfo {editor} {edited by\ \bibinfo
  {editor} {\bibfnamefont {J~\textit{et al}}\ \bibnamefont {Abad}}}\ (\bibinfo
  {publisher} {World Scientific Singapore},\ \bibinfo {year} {1986})\ pp.\
  \bibinfo {pages} {130--188}\BibitemShut {NoStop}%
\bibitem [{\citenamefont {Agrawal}\ \emph {et~al.}(2022)\citenamefont
  {Agrawal}, \citenamefont {Pandey}, \citenamefont {Kylhammar}, \citenamefont
  {Dev},\ and\ \citenamefont {Mitra}}]{agrawal2022memc}%
  \BibitemOpen
  \bibfield  {author} {\bibinfo {author} {\bibfnamefont {Vipin}\ \bibnamefont
  {Agrawal}}, \bibinfo {author} {\bibfnamefont {Vikash}\ \bibnamefont
  {Pandey}}, \bibinfo {author} {\bibfnamefont {Hanna}\ \bibnamefont
  {Kylhammar}}, \bibinfo {author} {\bibfnamefont {Apurba}\ \bibnamefont {Dev}},
  \ and\ \bibinfo {author} {\bibfnamefont {Dhrubaditya}\ \bibnamefont
  {Mitra}},\ }\bibfield  {title} {\enquote {\bibinfo {title} {Memc: A package
  for monte carlo simulations of spherical shells},}\ }\href {\doibase
  10.21105/joss.04305} {\bibfield  {journal} {\bibinfo  {journal} {Journal of
  Open Source Software}\ }\textbf {\bibinfo {volume} {7}},\ \bibinfo {pages}
  {4305} (\bibinfo {year} {2022})}\BibitemShut {NoStop}%
\bibitem [{\citenamefont {Ramaswamy}\ \emph {et~al.}(2000)\citenamefont
  {Ramaswamy}, \citenamefont {Toner},\ and\ \citenamefont
  {Prost}}]{ramaswamy2000nonequilibrium}%
  \BibitemOpen
  \bibfield  {author} {\bibinfo {author} {\bibfnamefont {Sriram}\ \bibnamefont
  {Ramaswamy}}, \bibinfo {author} {\bibfnamefont {John}\ \bibnamefont {Toner}},
  \ and\ \bibinfo {author} {\bibfnamefont {Jacques}\ \bibnamefont {Prost}},\
  }\bibfield  {title} {\enquote {\bibinfo {title} {Nonequilibrium fluctuations,
  traveling waves, and instabilities in active membranes},}\ }\href@noop {}
  {\bibfield  {journal} {\bibinfo  {journal} {Physical review letters}\
  }\textbf {\bibinfo {volume} {84}},\ \bibinfo {pages} {3494} (\bibinfo {year}
  {2000})}\BibitemShut {NoStop}%
\bibitem [{\citenamefont {Rao}\ and\ \citenamefont
  {R.C.}(2001)}]{rao2001active}%
  \BibitemOpen
  \bibfield  {author} {\bibinfo {author} {\bibfnamefont {Madan}\ \bibnamefont
  {Rao}}\ and\ \bibinfo {author} {\bibfnamefont {Sarasij}\ \bibnamefont
  {R.C.}},\ }\bibfield  {title} {\enquote {\bibinfo {title} {Active fusion and
  fission processes on a fluid membrane},}\ }\href@noop {} {\bibfield
  {journal} {\bibinfo  {journal} {Physical review letters}\ }\textbf {\bibinfo
  {volume} {87}},\ \bibinfo {pages} {128101} (\bibinfo {year}
  {2001})}\BibitemShut {NoStop}%
\bibitem [{\citenamefont {Loubet}\ \emph {et~al.}(2012)\citenamefont {Loubet},
  \citenamefont {Seifert},\ and\ \citenamefont
  {Lomholt}}]{loubet2012effective}%
  \BibitemOpen
  \bibfield  {author} {\bibinfo {author} {\bibfnamefont {Bastien}\ \bibnamefont
  {Loubet}}, \bibinfo {author} {\bibfnamefont {Udo}\ \bibnamefont {Seifert}}, \
  and\ \bibinfo {author} {\bibfnamefont {Michael~Andersen}\ \bibnamefont
  {Lomholt}},\ }\bibfield  {title} {\enquote {\bibinfo {title} {Effective
  tension and fluctuations in active membranes},}\ }\href@noop {} {\bibfield
  {journal} {\bibinfo  {journal} {Physical Review E}\ }\textbf {\bibinfo
  {volume} {85}},\ \bibinfo {pages} {031913} (\bibinfo {year}
  {2012})}\BibitemShut {NoStop}%
\bibitem [{\citenamefont {Maitra}\ \emph {et~al.}(2014)\citenamefont {Maitra},
  \citenamefont {Srivastava}, \citenamefont {Rao},\ and\ \citenamefont
  {Ramaswamy}}]{maitra2014activating}%
  \BibitemOpen
  \bibfield  {author} {\bibinfo {author} {\bibfnamefont {Ananyo}\ \bibnamefont
  {Maitra}}, \bibinfo {author} {\bibfnamefont {Pragya}\ \bibnamefont
  {Srivastava}}, \bibinfo {author} {\bibfnamefont {Madan}\ \bibnamefont {Rao}},
  \ and\ \bibinfo {author} {\bibfnamefont {Sriram}\ \bibnamefont {Ramaswamy}},\
  }\bibfield  {title} {\enquote {\bibinfo {title} {Activating membranes},}\
  }\href@noop {} {\bibfield  {journal} {\bibinfo  {journal} {Physical review
  letters}\ }\textbf {\bibinfo {volume} {112}},\ \bibinfo {pages} {258101}
  (\bibinfo {year} {2014})}\BibitemShut {NoStop}%
\bibitem [{\citenamefont {Hawkins}\ and\ \citenamefont
  {Liverpool}(2014)}]{hawkins2014stress}%
  \BibitemOpen
  \bibfield  {author} {\bibinfo {author} {\bibfnamefont {Rhoda~J}\ \bibnamefont
  {Hawkins}}\ and\ \bibinfo {author} {\bibfnamefont {Tanniemola~B}\
  \bibnamefont {Liverpool}},\ }\bibfield  {title} {\enquote {\bibinfo {title}
  {Stress reorganization and response in active solids},}\ }\href@noop {}
  {\bibfield  {journal} {\bibinfo  {journal} {Physical review letters}\
  }\textbf {\bibinfo {volume} {113}},\ \bibinfo {pages} {028102} (\bibinfo
  {year} {2014})}\BibitemShut {NoStop}%
\bibitem [{\citenamefont {Yin}\ \emph {et~al.}(2021)\citenamefont {Yin},
  \citenamefont {Li},\ and\ \citenamefont {Feng}}]{yin2021bio}%
  \BibitemOpen
  \bibfield  {author} {\bibinfo {author} {\bibfnamefont {Sifan}\ \bibnamefont
  {Yin}}, \bibinfo {author} {\bibfnamefont {Bo}~\bibnamefont {Li}}, \ and\
  \bibinfo {author} {\bibfnamefont {Xi-Qiao}\ \bibnamefont {Feng}},\ }\bibfield
   {title} {\enquote {\bibinfo {title} {Bio-chemo-mechanical theory of active
  shells},}\ }\href@noop {} {\bibfield  {journal} {\bibinfo  {journal} {Journal
  of the Mechanics and Physics of Solids}\ }\textbf {\bibinfo {volume} {152}},\
  \bibinfo {pages} {104419} (\bibinfo {year} {2021})}\BibitemShut {NoStop}%
\bibitem [{\citenamefont {Goriely}(2017)}]{goriely2017five}%
  \BibitemOpen
  \bibfield  {author} {\bibinfo {author} {\bibfnamefont {A}~\bibnamefont
  {Goriely}},\ }\bibfield  {title} {\enquote {\bibinfo {title} {Five ways to
  model active processes in elastic solids: active forces, active stresses,
  active strains, active fibers, and active metrics},}\ }\href@noop {}
  {\bibfield  {journal} {\bibinfo  {journal} {Mech. Res. Commun}\ }\textbf
  {\bibinfo {volume} {93}},\ \bibinfo {pages} {75--79} (\bibinfo {year}
  {2017})}\BibitemShut {NoStop}%
\bibitem [{\citenamefont {K{\"u}nsch}(1984)}]{kunsch1984non}%
  \BibitemOpen
  \bibfield  {author} {\bibinfo {author} {\bibfnamefont {H}~\bibnamefont
  {K{\"u}nsch}},\ }\bibfield  {title} {\enquote {\bibinfo {title} {Non
  reversible stationary measures for infinite interacting particle systems},}\
  }\href@noop {} {\bibfield  {journal} {\bibinfo  {journal} {Zeitschrift
  f{\"u}r Wahrscheinlichkeitstheorie und Verwandte Gebiete}\ }\textbf {\bibinfo
  {volume} {66}},\ \bibinfo {pages} {407--424} (\bibinfo {year}
  {1984})}\BibitemShut {NoStop}%
\bibitem [{\citenamefont {Godreche}\ and\ \citenamefont
  {Bray}(2009)}]{godreche2009nonequilibrium}%
  \BibitemOpen
  \bibfield  {author} {\bibinfo {author} {\bibfnamefont {Claude}\ \bibnamefont
  {Godreche}}\ and\ \bibinfo {author} {\bibfnamefont {Alan~J}\ \bibnamefont
  {Bray}},\ }\bibfield  {title} {\enquote {\bibinfo {title} {Nonequilibrium
  stationary states and phase transitions in directed ising models},}\
  }\href@noop {} {\bibfield  {journal} {\bibinfo  {journal} {Journal of
  Statistical Mechanics: Theory and Experiment}\ }\textbf {\bibinfo {volume}
  {2009}},\ \bibinfo {pages} {P12016} (\bibinfo {year} {2009})}\BibitemShut
  {NoStop}%
\bibitem [{\citenamefont {Godr{\`e}che}(2013)}]{godreche2013rates}%
  \BibitemOpen
  \bibfield  {author} {\bibinfo {author} {\bibfnamefont {Claude}\ \bibnamefont
  {Godr{\`e}che}},\ }\bibfield  {title} {\enquote {\bibinfo {title} {Rates for
  irreversible gibbsian ising models},}\ }\href@noop {} {\bibfield  {journal}
  {\bibinfo  {journal} {Journal of Statistical Mechanics: Theory and
  Experiment}\ }\textbf {\bibinfo {volume} {2013}},\ \bibinfo {pages} {P05011}
  (\bibinfo {year} {2013})}\BibitemShut {NoStop}%
\bibitem [{\citenamefont {Prost}\ and\ \citenamefont
  {Bruinsma}(1996)}]{prost1996shape}%
  \BibitemOpen
  \bibfield  {author} {\bibinfo {author} {\bibfnamefont {J}~\bibnamefont
  {Prost}}\ and\ \bibinfo {author} {\bibfnamefont {R}~\bibnamefont
  {Bruinsma}},\ }\bibfield  {title} {\enquote {\bibinfo {title} {Shape
  fluctuations of active membranes},}\ }\href@noop {} {\bibfield  {journal}
  {\bibinfo  {journal} {EPL (Europhysics Letters)}\ }\textbf {\bibinfo {volume}
  {33}},\ \bibinfo {pages} {321} (\bibinfo {year} {1996})}\BibitemShut
  {NoStop}%
\bibitem [{Note2()}]{Note2}%
  \BibitemOpen
  \bibinfo {note} {A. Fragkiadoulakis, S.K. Manikandan, and D. Mitra,
  unpublished.}\BibitemShut {Stop}%
\bibitem [{\citenamefont {Phillips}\ \emph {et~al.}(2012)\citenamefont
  {Phillips}, \citenamefont {Kondev}, \citenamefont {Theriot}, \citenamefont
  {Garcia},\ and\ \citenamefont {Orme}}]{phillips2012physical}%
  \BibitemOpen
  \bibfield  {author} {\bibinfo {author} {\bibfnamefont {Rob}\ \bibnamefont
  {Phillips}}, \bibinfo {author} {\bibfnamefont {Jane}\ \bibnamefont {Kondev}},
  \bibinfo {author} {\bibfnamefont {Julie}\ \bibnamefont {Theriot}}, \bibinfo
  {author} {\bibfnamefont {Hernan~G}\ \bibnamefont {Garcia}}, \ and\ \bibinfo
  {author} {\bibfnamefont {Nigel}\ \bibnamefont {Orme}},\ }\href@noop {} {\emph
  {\bibinfo {title} {Physical biology of the cell}}}\ (\bibinfo  {publisher}
  {Garland Science},\ \bibinfo {year} {2012})\BibitemShut {NoStop}%
\bibitem [{\citenamefont {Meyer}\ \emph {et~al.}(2003)\citenamefont {Meyer},
  \citenamefont {Desbrun}, \citenamefont {Schr{\"o}der},\ and\ \citenamefont
  {Barr}}]{meyer2003discrete}%
  \BibitemOpen
  \bibfield  {author} {\bibinfo {author} {\bibfnamefont {Mark}\ \bibnamefont
  {Meyer}}, \bibinfo {author} {\bibfnamefont {Mathieu}\ \bibnamefont
  {Desbrun}}, \bibinfo {author} {\bibfnamefont {Peter}\ \bibnamefont
  {Schr{\"o}der}}, \ and\ \bibinfo {author} {\bibfnamefont {Alan~H}\
  \bibnamefont {Barr}},\ }\bibfield  {title} {\enquote {\bibinfo {title}
  {Discrete differential-geometry operators for triangulated 2-manifolds},}\
  }in\ \href@noop {} {\emph {\bibinfo {booktitle} {Visualization and
  mathematics III}}}\ (\bibinfo  {publisher} {Springer},\ \bibinfo {year}
  {2003})\ pp.\ \bibinfo {pages} {35--57}\BibitemShut {NoStop}%
\bibitem [{\citenamefont {Hege}\ and\ \citenamefont
  {Polthier}(2003)}]{hege2003visualization}%
  \BibitemOpen
  \bibfield  {author} {\bibinfo {author} {\bibfnamefont {Hans-Christian}\
  \bibnamefont {Hege}}\ and\ \bibinfo {author} {\bibfnamefont {Konrad}\
  \bibnamefont {Polthier}},\ }\href@noop {} {\emph {\bibinfo {title}
  {Visualization and mathematics III}}}\ (\bibinfo  {publisher} {Springer
  Science \& Business Media},\ \bibinfo {year} {2003})\BibitemShut {NoStop}%
\bibitem [{\citenamefont {Caroli}\ \emph {et~al.}(2010)\citenamefont {Caroli},
  \citenamefont {de~Castro}, \citenamefont {Loriot}, \citenamefont {Rouiller},
  \citenamefont {Teillaud},\ and\ \citenamefont {Wormser}}]{caroli2010robust}%
  \BibitemOpen
  \bibfield  {author} {\bibinfo {author} {\bibfnamefont {Manuel}\ \bibnamefont
  {Caroli}}, \bibinfo {author} {\bibfnamefont {Pedro~MM}\ \bibnamefont
  {de~Castro}}, \bibinfo {author} {\bibfnamefont {S{\'e}bastien}\ \bibnamefont
  {Loriot}}, \bibinfo {author} {\bibfnamefont {Olivier}\ \bibnamefont
  {Rouiller}}, \bibinfo {author} {\bibfnamefont {Monique}\ \bibnamefont
  {Teillaud}}, \ and\ \bibinfo {author} {\bibfnamefont {Camille}\ \bibnamefont
  {Wormser}},\ }\bibfield  {title} {\enquote {\bibinfo {title} {Robust and
  efficient delaunay triangulations of points on or close to a sphere},}\ }in\
  \href@noop {} {\emph {\bibinfo {booktitle} {International Symposium on
  Experimental Algorithms}}}\ (\bibinfo {organization} {Springer},\ \bibinfo
  {year} {2010})\ pp.\ \bibinfo {pages} {462--473}\BibitemShut {NoStop}%
\bibitem [{A simple and fast mesh generator()}]{meshzoo}%
  \BibitemOpen
  A simple and fast mesh generator,\ \href@noop {} {}\bibinfo {howpublished}
  {\url{https://github.com/meshpro/meshzoo}}\BibitemShut {NoStop}%
\bibitem [{\citenamefont {Hunter}(2007)}]{Hunter:2007}%
  \BibitemOpen
  \bibfield  {author} {\bibinfo {author} {\bibfnamefont {J.~D.}\ \bibnamefont
  {Hunter}},\ }\bibfield  {title} {\enquote {\bibinfo {title} {Matplotlib: A 2d
  graphics environment},}\ }\href {\doibase 10.1109/MCSE.2007.55} {\bibfield
  {journal} {\bibinfo  {journal} {Computing in Science \& Engineering}\
  }\textbf {\bibinfo {volume} {9}},\ \bibinfo {pages} {90--95} (\bibinfo {year}
  {2007})}\BibitemShut {NoStop}%
\bibitem [{\citenamefont {Childs}\ \emph {et~al.}(2012)\citenamefont {Childs},
  \citenamefont {Brugger}, \citenamefont {Whitlock}, \citenamefont {Meredith},
  \citenamefont {Ahern}, \citenamefont {Pugmire}, \citenamefont {Biagas},
  \citenamefont {Miller}, \citenamefont {Harrison}, \citenamefont {Weber},
  \citenamefont {Krishnan}, \citenamefont {Fogal}, \citenamefont {Sanderson},
  \citenamefont {Garth}, \citenamefont {Bethel}, \citenamefont {Camp},
  \citenamefont {R\"{u}bel}, \citenamefont {Durant}, \citenamefont {Favre},\
  and\ \citenamefont {Navr\'{a}til}}]{HPV:VisIt}%
  \BibitemOpen
  \bibfield  {author} {\bibinfo {author} {\bibfnamefont {Hank}\ \bibnamefont
  {Childs}}, \bibinfo {author} {\bibfnamefont {Eric}\ \bibnamefont {Brugger}},
  \bibinfo {author} {\bibfnamefont {Brad}\ \bibnamefont {Whitlock}}, \bibinfo
  {author} {\bibfnamefont {Jeremy}\ \bibnamefont {Meredith}}, \bibinfo {author}
  {\bibfnamefont {Sean}\ \bibnamefont {Ahern}}, \bibinfo {author}
  {\bibfnamefont {David}\ \bibnamefont {Pugmire}}, \bibinfo {author}
  {\bibfnamefont {Kathleen}\ \bibnamefont {Biagas}}, \bibinfo {author}
  {\bibfnamefont {Mark}\ \bibnamefont {Miller}}, \bibinfo {author}
  {\bibfnamefont {Cyrus}\ \bibnamefont {Harrison}}, \bibinfo {author}
  {\bibfnamefont {Gunther~H.}\ \bibnamefont {Weber}}, \bibinfo {author}
  {\bibfnamefont {Hari}\ \bibnamefont {Krishnan}}, \bibinfo {author}
  {\bibfnamefont {Thomas}\ \bibnamefont {Fogal}}, \bibinfo {author}
  {\bibfnamefont {Allen}\ \bibnamefont {Sanderson}}, \bibinfo {author}
  {\bibfnamefont {Christoph}\ \bibnamefont {Garth}}, \bibinfo {author}
  {\bibfnamefont {E.~Wes}\ \bibnamefont {Bethel}}, \bibinfo {author}
  {\bibfnamefont {David}\ \bibnamefont {Camp}}, \bibinfo {author}
  {\bibfnamefont {Oliver}\ \bibnamefont {R\"{u}bel}}, \bibinfo {author}
  {\bibfnamefont {Marc}\ \bibnamefont {Durant}}, \bibinfo {author}
  {\bibfnamefont {Jean~M.}\ \bibnamefont {Favre}}, \ and\ \bibinfo {author}
  {\bibfnamefont {Paul}\ \bibnamefont {Navr\'{a}til}},\ }\bibfield  {title}
  {\enquote {\bibinfo {title} {{VisIt: An End-User Tool For Visualizing and
  Analyzing Very Large Data}},}\ }in\ \href@noop {} {\emph {\bibinfo
  {booktitle} {{High Performance Visualization--Enabling Extreme-Scale
  Scientific Insight}}}}\ (\bibinfo {year} {2012})\ pp.\ \bibinfo {pages}
  {357--372}\BibitemShut {NoStop}%
\end{thebibliography}
%
\end{document}